\documentclass[preprint,12pt]{aastex}               
\usepackage{epsfig}

\def\beq{\begin{equation}}
\def\eeq{\end{equation}}

\shortauthors{D.~A.~Uzdensky}

\begin{document}

\title{Force-Free Magnetosphere of an Accretion Disk --- Black Hole
System. II. Kerr Geometry}
\author{Dmitri A. Uzdensky
\thanks{Currently at Princeton University.}}
\affil{Kavli Institute for Theoretical Physics, University of California} 
\affil{Santa Barbara, CA 93106}
\email{uzdensky@kitp.ucsb.edu}
\date{October 27, 2004}

\begin{abstract}
We consider a stationary axisymmetric force-free degenerate 
magnetosphere of a rotating Kerr black hole surrounded by a
thin Keplerian infinitely-conducting accretion disk. We focus
on the closed-field geometry characterize by a direct magnetic 
coupling between the disk and the hole's event horizon. We first 
argue that the hole's rotation necessarily limits the radial extent 
of the force-free link on the disk surface: the faster the hole rotates, 
the smaller the magnetically-connected inner part of the disk has to be. 
We then show that this is indeed the case by solving numerically the 
Grad--Shafranov equation---the main differential equation describing 
the structure of the magnetosphere. An important element in our approach 
is the use of the regularity condition at the inner light cylinder to 
fix the poloidal current as a function of the poloidal magnetic flux. 
As an outcome of our computations, we are able to chart out the maximum 
allowable size of the portion of the disk that is magnetically connected 
to the hole as a function of the black hole spin. We also calculate the 
angular momentum and energy transfer between the hole and the disk that 
takes place via the direct magnetic link. We find that both of these 
quantities grow rapidly and that their deposition becomes highly 
concentrated near the inner edge of the disk as the black hole spin 
is increased.

\end{abstract}



\section{Introduction}
\label{sec-intro}

This paper is devoted to the subject of magnetic interaction between 
a rotating black hole and an accretion disk around it---a topic that
has enjoyed a lot of attention among researchers in recent years.
Magnetic fields are believed to play an important role in the dynamics 
of accreting black hole systems (e.g., Begelman, Blandford, \& Rees~1984; 
Krolik~1999b; Punsly~2001). In particular, they can be very effective in 
transporting angular momentum and the associated rotational energy of 
either the hole or the disk. 

Where and how this transport takes place and to what observational 
consequences it can lead, is partly determined by the global geometry 
of the magnetic field lines. Conceptually, one can think of two basic 
types of geometry. The first type is the open-field configuration
shown schematically in Figure~\ref{fig-geometry-open}. The main 
topological feature here is that there is no direct magnetic link 
between the hole and the disk. Instead, all the field lines are open 
and extend to infinity. Historically, this configuration was the first 
to have been considered, and it has been studied very extensively during 
the past three decades (see, e.g., Lovelace~1976; Blandford~1976; Blandford 
\& Znajek~1977, hereafter BZ77; MacDonald \& Thorne~1982, hereafter MT82; 
Phinney~1983; Macdonald~1984; Thorne, Price, \& Macdonald~1986; 
Punsly 1989, 2001, 2003, 2004; Punsly \& Coroniti 1990; Beskin \& Par'ev 1993; 
Beskin~1997; Ghosh \& Abramowicz 1997; Beskin \& Kuznetsova 2000; 
Komissarov~2001, 2002b, 2004a). The reason for this popularity is that 
this configuration is related to the famous Blandford--Znajek mechanism
(BZ77) now widely regarded as the primary process powering jets in active 
galactic nuclei (AGN) and micro-quasars. As Blandford and Znajek showed, 
a large-scale, ordered open magnetic field can extract the rotational 
energy from a spinning black hole and transport it to large distances
via Poynting flux (a similar process works along the field lines 
connected to the disk). 

The second type of magnetic field geometry is the closed-field configuration, 
shown in Figure~\ref{fig-geometry-closed}. Although it has been occasionally
discussed in the literature before the last decade (e.g., Zeldovich \& 
Schwartzman, quoted in Thorne~1974; MT82; Thorne~et~al. 1986; Nitta, 
Takahashi, \&  Tomimatsu~1991; Hirotani~1999), it is only in the last 
five years that it has attracted serious scientific attention 
(e.g., Blandford~1999, 2000, 2002; Gruzinov~1999; van~Putten 1999; 
van~Putten \& Levinson 2003; 
Li~2000, 2001, 2002a, 2002b, 2004; Wang, Xiao, \& Lei 2002; 
Wang, Lei, \& Ma 2003a; Wang et al. 2003b, 2004).
The basic topological structure of magnetic field in this configuration
is very different from that of the open-field configuration.
The field lines are closed and directly connect the black hole 
to the disk. In this so-called Magnetically-Coupled configuration 
(Wang et al. 2002), the energy and angular momentum are not taken away 
to infinity, but instead are exchanged between the hole and the disk by
the magnetic field. Therefore, magnetic coupling, together with the 
accretion process, controls the spin evolution and the spin equilibrium 
of the black hole (Wang et al. 2002, 2003a). In addition, the 
rotational energy of the hole can be magnetically extracted (just like 
in the Blandford--Znajek process) and deposited onto the disk leading to
a change in the disk energy-dissipation profile and hence its observable 
spectral characteristics (Gammie 1999; Li 2000, 2001, 2002a, 2002b, 2004;
Wang et al. 2003a,b). 
Finally, if the rotating field lines are strongly twisted and become unstable 
to a non-axisymmetric kink-like instability, a strong variability of
the energy release may result, which would be a possible explanation
for quasi-periodic oscillations (QPOs) in micro-quasar systems (e.g.,
Gruzinov~1999; Wang et al.~2004). All these phenomena make the closed-field 
configuration astrophysically very interesting.

Most of the work that has been done on studying the magnetic field
structure around an accreting Kerr black hole, including the seminal 
paper by Blandford \& Znajek (BZ77), has been performed under the 
assumption that the magnetosphere above the thin disk is ideally 
conducting and force-free. Then, if one also assumes that the system 
is stationary and axisymmetric, the magnetic field is governed by the 
general-relativistic version of the force-free Grad--Shafranov equation 
(e.g., MT82; for the full-MHD generalization of this equation see 
Nitta~et~al. 1991; Beskin \& Par'ev 1993; Beskin~1997). Since this 
is a rather nontrivial nonlinear partial differential equation (PDE) 
with singular surfaces and free functions, it is generally not tractable 
analytically, except in some special simple cases, such as the slow-rotation 
limit (BZ77). However, over the past 20 years, a number of force-free
solutions for the magnetosphere have been obtained numerically, either 
by solving the Grad--Shafranov equation itself (MacDonald~1984; Fendt~1997) 
or as an asymptotic steady state of force-free degenerate electrodynamics 
(FFDE) evolution (Komissarov~2001, 2002b, 2004a). Until now, most of these 
studies have been done in the context of the open-field configuration, 
primarily because of its relevance to the jet problem. 

In contrast, most of the work on closed-field configurations
has been limited to analytic and semi-analytic studies of the 
effects that magnetic link has on the disk radiation profile 
and on the spin evolution of the black hole. The structure of
the magnetosphere has not in fact been computed self-consistently.
These studies have just assumed the existence of the link and 
made some simplified assumptions about the field distribution 
on the horizon.

The only exception to this deficiency is the recent work by Uzdensky 
(2004) where a force-free magnetosphere linking a Keplerian disk to 
a Schwarzschild black hole has been numerically computed for the 
first time. 

In the present paper, we make the next logical step by extending this
previous work to the more general case of a rapidly rotating Kerr black 
hole. This is indeed the most important case, not only because real 
astrophysical black holes are believed to be rotating, but also because 
the nonlinear terms in the Grad--Shafranov equation, especially the 
toroidal field pressure, become large in this case. As a result, even 
the existence of closed-field solutions is not guaranteed. And indeed, 
one of the main goals of our present study is to determine the conditions 
for existence of such solutions in Kerr geometry. In other words, we aim 
at determining the limitations that the rotation of the black hole imposes 
on the direct magnetic link between the hole and the disk. In addition, by 
computing the global magnetic field structure, we will be able to study the 
effect of the black hole rotation on the magnetic field distribution on the 
horizon, the poloidal electric current as a function of poloidal magnetic 
flux, and the location of the inner light cylinder, as well as such 
astrophysically-important processes as angular momentum and energy 
transfer between the hole and the disk.

In order to achieve the goal of obtaining numerical solutions of the
force-free Grad--Shafranov equation in Kerr goemetry, we first analize 
the mathematical structure of this equation. In particular, we pay
special attention to its singular surfaces (the event horizon and
the light cylinder) and the corresponding regularity conditions.
Thus, we use the light-cylinder regularity condition to determine
the poloidal current as a function of poloidal magnetic flux, similar
to way it was done by Contopoulos, Kazanas, \& Fendt (1999) for the
case of the pulsar magnetosphere (see also Beskin \& Kuznetsova 2000;
Uzdensky~2003; Uzdensky~2004). The event-horizon regularity condition,
also known as Znajek's (1977) horizon boundary condition, is then
used to determine the poloidal flux distribution on the horizon. Thus,
one does not have the freedom to arbitrarily specify any extra boundary 
conditions at the horizon, and hence there is no problem with causality,
in line with the reasoning presented by Beskin \& Kuznetsova (2000) and
by Komissarov (2002b, 2004a) (see also Levinson~2004).

Finally, although in this paper we deal exclusively with large-scale,
ordered magnetic fields, we acknowledge the difficulty in justifying
the existence of such fields around accreting black holes (e.g., Livio, 
Ogilvie, \& Pringle 1999), especially in the closed-field configuration.
Also, as recent numerical simulations (e.g., Hawley \& Krolik 2001;
Hirose et al.~2004), there may be a significant deposition of energy 
and angular momentum at the inner edge of the disk due to small-scale, 
intermittent magnetic fields connecting the disk to the plunging region 
(see also Krolik~1999a; Agol \& Krolik~2000).

The paper is organized as follows. \S~\ref{sec-equations} describes 
the mathematical formalism of force-free axisymmetric stationary 
magnetospheres in Kerr geometry. In particular, in \S~\ref{subsec-kerr} 
we introduce the Kerr metric tensor in Boyer--Lindquist coordinates and 
list several general geometric relationships for the future use.
In \S~\ref{subsec-GS-eqn} we consider steady-state, axisymmetric, 
degenerate electro-magnetic fields and then discuss the force-free 
condition and the Grad--Shafranov equation. In \S~\ref{subsec-EH} 
we consider the black hole's event horizon as a singular surface 
of this equation and discuss the associated regularity condition, 
which is also known as Znajek's horizon boundary condition.
In \S~\ref{sec-idea} we present a simple but robust physical 
argument that demonstrates that a force-free magnetic link 
between a rotating black hole and the disk cannot extend to 
arbitrarily large distances on the disk, we also argue that 
the maximal radial extent of the magnetic link should scale 
inversely with the black hole's rotation rate in the slow-rotation 
limit. We confirm these propositions in \S~\ref{sec-numerical}, 
where we present our numerical solutions of the grad--Shafranov 
equation. Then, in \S~\ref{sec-implications} we discuss the 
magnetically-mediated angular-momentum and energy exchange 
between the hole and the disk. We then close by summarizing 
our findings in \S~\ref{sec-conclusions}.


\section{Axisymmetric force-free magnetosphere in Kerr geometry ---
basic equations}
\label{sec-equations}

\subsection{Kerr geometry --- mathematical preliminaries}
\label{subsec-kerr}

In this paper we employ Boyer--Lindquist coordinates 
($t,r,\theta,\phi$) in Kerr geometry. The metric of the 
four-dimensional space-time can be written in these coordinates
as
\beq
ds^2 = (g_{\phi\phi}\omega^2 - \alpha^2) dt^2 - 
2\omega g_{\phi\phi} d\phi dt + g_{rr} dr^2 +
g_{\theta\theta} d\theta^2 + g_{\phi\phi}d\phi^2 \, ,
\label{eq-metric}
\eeq
with the components of the metric tensor given by
\begin{eqnarray}
\alpha &=& {\rho\over\Sigma} \sqrt{\Delta} \, ,
\label{eq-alpha}  \\
\omega &=& {2aMr\over{\Sigma^2}} \, ,
\label{eq-beta}   \\
g_{rr} &=& {\rho^2\over\Delta}, \qquad
g_{\theta\theta}=\rho^2, \qquad
g_{\phi\phi} = \varpi^2 \, ,
\label{eq-metric-tensor} 
\end{eqnarray}
where 
\begin{eqnarray}
\rho^2 &\equiv& r^2+a^2 \cos^2\theta \, , 
\label{eq-rho}  \\
\Delta &\equiv& r^2+a^2-2Mr \, , 
\label{eq-Delta} \\
\Sigma^2 &\equiv& (r^2+a^2)^2 - a^2\Delta\sin^2\theta  \, ,
\label{eq-Sigma} \\
\varpi &\equiv& {\Sigma\over\rho} \sin\theta \, .
\label{eq-varpi}
\end{eqnarray}

Here, $M$ and $a\in [0;M]$ are the mass and the spin parameter 
(specific angular momentum) of the central black hole, respectively. 
(Throughout this paper we use geometric units, i.e., we set both the 
gravitational constant $G$ and the speed of light $c$ to~1). 

In order to describe the electromagnetic processes around a black hole,
we use the 3+1 split of the laws of electrodynamics introduced by MT82 
(see also Thorne~et~al. 1986). In this formalism, one splits the 
four-dimensional 
spacetime into the global time $t$ and the absolute three-dimensional 
curved space, the geometry of which is described by a three-dimensional 
(3D) metric tensor with components given by equation~(\ref{eq-metric-tensor}). 
The electromagnetic field is represented by the electric and magnetic field 
3-vectors ${\bf E}$ and ${\bf B}$ measured by local zero-angular-momentum 
observers (ZAMOs; see Thorne~et~al. 1986). In order to describe these vectors, we will 
use both the coordinate basis $\{ \partial_i \} = \{{\bf e}_i\}$ and the 
orthonormal basis $\{{\bf e}_{\hat{i}} \}$ [where the Roman index $i$ 
runs through the three spatial coordinates $(r,\theta,\phi)$]. Because 
the spatial 3D metric tensor $g_{ij}$ is diagonal, these two bases 
are related via
\beq
{\bf e}_i = \sqrt{g_{ii}} \, {\bf e}_{\hat{i}} \, ,
\qquad i=r,\theta,\phi \, 
\label{eq-basis-1} 
\eeq
(note: there is no summation over $i$ in this expression!).
In particular, in the Boyer--Lindquist coordinates in Kerr 
geometry, we have
\beq
{\bf e}_r = {\rho\over\sqrt{\Delta}}\, {\bf e}_{\hat{r}} \, , \qquad
{\bf e}_\theta = \rho {\bf e}_{\hat{\theta}} \, , \qquad
{\bf e}_\phi = \varpi \, {\bf e}_{\hat{\phi}} \, .
\label{eq-basis-2}
\eeq

We shall also need the following mathematical expressions:
the 3-gradient of a scalar function $f({\bf x})=f(r,\theta,\phi)$ 
in the Boyer--Lindquist coordinates is 
\begin{eqnarray}
\nabla f &=& \sum\limits_i \, 
g_{ii}^{-1/2} (\partial_i f) {\bf e}_{\hat{i}} \nonumber \\
&=& {\sqrt{\Delta}\over\rho}\, (\partial_r f)\, {\bf e}_{\hat r} +
{1\over\rho}\, (\partial_\theta f)\, {\bf e}_{\hat \theta} +
{1\over\varpi}\, (\partial_\phi f)\, {\bf e}_{\hat \phi} \, ,
\label{eq-gradient} 
\end{eqnarray}
and its square is 
\begin{eqnarray}
|\nabla f|^2 &=& \sum\limits_i g_{ii}^{-1} (\partial_i f)^2 \nonumber \\
&=& {\Delta\over{\rho^2}} (\partial_r f)^2 +
{1\over{\rho^2}} (\partial_\theta f)^2 +
{1\over{\varpi^2}} (\partial_\phi f)^2 \, .
\label{eq-gradient-square} 
\end{eqnarray}

Finally, the 3-divergence of a 3-vector ${\bf A}$ can be written as
\beq
\nabla \cdot {\bf A} = A^i_{;i}=
A^i_{,i} + A^i \bigl(\ln\sqrt{|g|}\bigr)_{,i} =
{1\over{\sqrt{|g|}}}\, \biggl( \sqrt{|g|} A^i \biggr)_{,i} \ ,
\label{eq-divergence} 
\eeq
where $g$ is the determinant of the 3-D metric tensor:
\beq
\sqrt{|g|} = {{\rho^2\varpi}\over{\sqrt{\Delta}}} = 
{{\rho\Sigma\sin\theta}\over{\sqrt{\Delta}}} \, .
\label{eq-g}
\eeq


\subsection{Stationary axisymmetric ideal force-free magnetosphere 
in Kerr geometry}
\label{subsec-GS-eqn}

As mentioned above, in the 3+1 split formalism of MT82 a magnetosphere 
of a rotating Kerr black hole is described in terms of two spatial vector 
fields, ${\bf E}$ and ${\bf B}$.
Under the assumptions that the magnetosphere is: (1) stationary
($\partial_t =0$), (2) axisymmetric ($\partial_\phi=0$), and
(3) ideally-conducting, or degenerate (${\bf E\cdot B} =0$), these
two vector fields can be expressed in terms of three scalar functions,
$\Psi(r,\theta)$, $\Omega_F(r,\theta)$, and $I(r,\theta)$:
\beq
{\bf B} (r,\theta) = {\bf B}_{\rm pol} + {\bf B}_{\rm tor} \, ,
\eeq
where
\begin{eqnarray}
{\bf B}_{\rm pol} &=& \nabla\Psi\times\nabla\phi =
{1\over{\varpi\rho}}\, \Psi_\theta\, {\bf e}_{\hat{r}} -
{\sqrt{\Delta}\over{\varpi\rho}}\, \Psi_r\, {\bf e}_{\hat{\theta}}\, ,
\label{eq-Bpol} \\
{\bf B}_{\rm tor} &=& B_{\hat{\phi}} {\bf e}_{\hat{\phi}} = 
{I\over{\alpha\varpi}}\, {\bf e}_{\hat{\phi}} \, ,
\label{eq-Btor}
\end{eqnarray} 
and
\begin{equation}
{\bf E} (r,\theta) = {\bf E}_{\rm pol} =
-{{\delta\Omega}\over\alpha}\, \nabla\Psi\, , \qquad 
E_{\phi} = 0 \, , 
\label{eq-E}
\eeq
where
\beq
\delta\Omega \equiv \Omega_F - \omega \, .
\label{eq-DeltaOmega}
\eeq

Here, $\Psi(r,\theta)$ is the poloidal magnetic flux function, 
$\Omega_F=\Omega_F(\Psi)$ is the angular velocity of the magnetic 
field lines, and $I(r,\theta)$ is $(2/c)$ times the poloidal electric 
current flowing through the circular loop $r={\rm const}$, $\theta=
{\rm const}$. 
[Note that our definitions of $\Psi$ and $I$ differ from the ones 
adopted by MT82: $\Psi=\psi_{\rm MT82}/2\pi$, $I=-(2/c)I_{\rm MT82}$.]

Next, in this work we are interested in the case of a {\it force-free}
magnetosphere, i.e., a magnetosphere that is so tenuous that the 
electromagnetic forces completely dominate over all others, including
gravitational, pressure, and inertial forces. Even though this framework 
has been widely accepted as a primary tool in describing magnetospheres 
of black holes and radio-pulsars, its usefulness and validity near
the event horizon has been seriously challenged by Punsly (2001, 2003). 
However, according to the recent MHD simulations by Komissarov (2004b),
these worries seem to be unfounded. Therefore, we shall still employ 
the force-free approach in this paper. Correspondingly, we shall write 
the force-balance equation (in the ZAMO reference frame) as
\beq
\rho_e {\bf E} + {\bf j \times B} = 0 \, ,
\label{eq-force-free}
\eeq
where the ZAMO-measured electric charge density $\rho_e$ and 
electric current density ${\bf j}$ are related to ${\bf E}$ 
and ${\bf B}$ via Maxwell's equations (see MT82).

The toroidal component of the force-free equation immediately leads to 
\beq
I(r,\theta) = I(\Psi) \, ,
\label{eq-I=IofPsi}
\eeq
i.e., the poloidal electric current does not cross poloidal 
flux surfaces.

The poloidal component of equation~(\ref{eq-force-free}), upon using 
expressions (15)--(18), yields the so-called generally-relativistic 
force-free Grad--Shafranov equation --- the main equation that governs 
the system. In this paper we shall use as a starting point the form of 
this equation given in MT82 (i.e., eq. [6.4] of MT82 slightly modified 
to account for the change in the definition of~$\Psi$):
\begin{eqnarray}
\nabla \cdot \biggl( {\alpha\over{\varpi^2}}\, 
\bigl[ 1-{{\delta\Omega^2 \varpi^2}\over{\alpha^2}} \bigr] 
\nabla\Psi \biggr) &+& \nonumber \\
{{\delta\Omega}\over\alpha}\, {d\Omega_F\over{d\Psi}}\, (\nabla\Psi)^2 +
{1\over{\alpha\varpi^2}}\, II'(\Psi) &=& 0 \, .
\label{eq-GS-MT82}
\end{eqnarray}

This is a nonlinear 2nd-order elliptic partial differential equation (PDE); 
it determines $\Psi(r,\theta)$ provided that $\Omega_F(\Psi)$ and $I(\Psi)$ 
are known. We can rewrite this equation as follows:
\beq
LHS \equiv \alpha\varpi^2 \nabla \cdot \biggl({1\over{\alpha\varpi^2}}\, 
(\alpha^2 - \delta\Omega^2 \varpi^2) \nabla\Psi \biggr) =
RHS \equiv -II'(\Psi) - \delta\Omega \Omega_F'(\Psi) \, 
\varpi^2 (\nabla\Psi)^2 \, ,
\label{eq-GS}
\eeq
where a prime denotes the derivative with respect to~$\Psi$,
e.g., $I'(\Psi)=dI/d\Psi$.

Upon introducing the quantities 
\beq
D \equiv \alpha^2 - \delta\Omega^2 \varpi^2, 
\label{eq-D-def}
\eeq
and
\beq
Q(r,\theta) \equiv {\sqrt{|g|}\over{\alpha\varpi^2}} =
{{\rho\Sigma}\over{\varpi\Delta}} = {{\rho^2}\over{\Delta\sin\theta}}\, ,
\label{eq-Q-def} 
\eeq
and upon using identity~(\ref{eq-divergence}), the left-hand side
(LHS) of this equation can be written in a compact and convenient form
\beq
LHS = Q^{-1} [QD(\nabla\Psi)^i]_{,i} =
[D(\nabla\Psi)^i]_{,i} + D(\nabla\Psi)^i \partial_i \ln Q \, .
\label{eq-LHS}
\eeq

Using expression~(\ref{eq-gradient}), we get the Grad--Shafranov 
equation in the following final form:
\begin{eqnarray}
LHS &=& \partial_r \biggl( {{D\Delta}\over{\rho^2}}\, \Psi_r \biggr) +
\partial_\theta \biggl( {D\over{\rho^2}}\, \Psi_\theta \biggr) +
{D\over{\rho^2 Q}}\, \biggl( \Psi_r\Delta\partial_r Q +
\Psi_\theta \partial_\theta Q \biggr) \nonumber \\
&=& RHS \equiv -II'(\Psi) - \delta\Omega \Omega_F'(\Psi) \, 
\varpi^2 (\nabla\Psi)^2 \, .
\label{eq-GS-2}
\end{eqnarray}


\subsection{Regularity condition at the event horizon}
\label{subsec-EH}

From the Grad--Shafranov equation in the form~(\ref{eq-GS-2}) 
it is easy to see that, in general, this equation has two types of 
singular surfaces. One of them is the so-called {\it light cylinder}
(often called in the literature the velocity-of-light surface or simply 
the light surface) defined as a surface where $D=0$. We shall discuss
it in more detail later (see \S~\ref{subsec-LC}).

There is also another singular surface of the Grad--Shafranov equation: 
{\it the event horizon} defined as the surface where
\beq
\Delta = 0 = \alpha \, .
\label{eq-EH}
\eeq
This surface will be the main focus of this section.

As can be seen from equation~(\ref{eq-Delta}), the event 
horizon is a constant-$r$ surface, 
\beq
r(\theta) = r_H = M + \sqrt{M^2-a^2} = {\rm const} \, .
\label{eq-rH}
\eeq
In addition, the frame-dragging frequency $\omega$ 
defined by equation~(\ref{eq-beta}) is also constant 
on the horizon,
\beq
\omega(r=r_H,\theta) = \Omega_H = {a\over{2Mr_H}} = {\rm const} \, .
\label{eq-OmegaH}
\eeq
This constant is what is conventionally called the rotation rate 
of the Kerr black hole.

Because the horizon is surface of constant~$r$, one can immediately 
see that it is a singular surface of equation~(\ref{eq-GS-2}). This
is because the coefficient in front of the 2nd-order derivative in 
the direction normal to this surface (in this case, radial) vanishes, 
even though the coefficient in front of the 2nd derivative in the 
$\theta$-direction does not.

The fact that the event horizon is just a singular surface of the 
Grad--Shafranov equation is extremely important. It means that 
one cannot impose an independent boundary condition for the function 
$\Psi(r,\theta)$ at the horizon. One can only impose a {\it regularity 
condition} there (Beskin~1997; Komissarov~2002b, 2004a; Uzdensky~2004).
Mathematically, this condition means that there should be no 
logarithmic terms in the asymptotic expansion of $\Psi(r,\theta)$
near $r=r_H$ (see MT82). Physically, the regularity condition originates
from the requirement that freely-falling observers measure finite electric 
and magnetic fields near the horizon (see Thorne~et~al. 1986). Alternatively,
the event horizon regularity condition can be obtained from the
fast-magnetosonic critical condition in the limit in which plasma
density goes to zero and hence the inner fast magnetosonic surface
approaches the horizon (Beskin~1997; Beskin \& Kuznetsova 2000; 
Komissarov~2004a).
In the present paper, we will not repeat the rigorous derivation 
of this condition (we refer the reader to MT82 or Thorne~et~al. 1986). 
Instead, we just note that as a result of the regularity requirement, 
one expects both the 1st and 2nd radial derivatives of $\Psi$ to remain 
finite at the horizon. Therefore, when applying the Grad--Shafranov 
equation~(\ref{eq-GS-2}) at $r=r_H$, one can just simply set $\Delta=0$. 
Then, after some algebra, one gets:
\beq
I^2[\Psi_0(\theta)] = 
\biggl(\delta\Omega {\varpi\over\rho}\, 
{{d\Psi_0}\over{d\theta}} \biggr)^2 +
{\rm const} \, , \qquad r=r_H \, ,
\eeq
where 
\beq
\Psi_0(\theta) \equiv \Psi(r=r_H,\theta) \, .
\label{eq-Psi0}
\eeq

In the absence of a finite line-current along the axis $\theta=0$,
i.e., when $I(\theta=0)=0$, the integration constant is zero and
hence
\beq
I = \pm \delta\Omega\, {\varpi\over\rho}\, {d\Psi_0\over{d\theta}}\, ,
\qquad r=r_H \, .
\label{eq-I-EH}
\eeq

As for the choice of sign in this expression, it can be 
shown that the correct sign must be plus [remember that 
MT82 have minus sign because we define $I(\Psi)$ with an
opposite sign]; this comes from the requirement that 
Poynting flux measured by a ZAMO in the vicinity of the 
horizon is directed {\it towards} the black hole (e.g.,
Znajek~1977, 1978; BZ77; MT82). Thus, we have
\beq
I[\Psi_0(\theta)]=\delta\Omega\, {\varpi\over\rho}\, {d\Psi_0\over{d\theta}}=
{{2Mr_H\sin\theta}\over{\rho^2}}\, \delta\Omega\, {d\Psi_0\over{d\theta}}\, ,
\qquad r=r_H \, .
\label{eq-EH-bc}
\eeq

Equation~(\ref{eq-EH-bc}) was first derived by Znajek (1977) and is 
frequently referred to as the "Znajek's horizon boundary condition".
We stress, however, that, because the event horizon is a singular
surface of the Grad--Shafranov equation, one cannot really impose 
a boundary condition there; expression~(\ref{eq-EH-bc}) actually 
follows from the Grad--Shafranov equation itself under the condition 
that the solution be regular at~$r=r_H$.

It is interesting to note that, because not only the 2nd-, but 
also the 1st-order radial derivatives of $\Psi$ drop out of the 
Grad--Shafranov equation when $\Delta$ is set to zero, this equation
becomes an ordinary (as opposed to a partial) differential equation
at the horizon! This implies that the horizon poloidal magnetic flux 
distribution, $\Psi_0(\theta)$, is connected to the magnetosphere 
outside the horizon only through the functions $I(\Psi)$ and 
$\Omega_F(\Psi)$ and not through any radial derivatives. From 
the practical point of view, this fact means that equation~(\ref{eq-EH-bc}) 
can be viewed as a Dirichlet-type boundary condition that determines 
the function~$\Psi_0(\theta)$  once both $I(\Psi)$ and $\Omega_F(\Psi)$ 
are given. It is important to emphasize that we really have only one 
relationship on the horizon--- equation~(\ref{eq-EH-bc}) --- between 
three functions [$\Psi_0(\theta)$, $I(\Psi)$, and $\Omega_F(\Psi)$] 
and hence one needs to find some other conditions, set somewhere else, 
to fix $I(\Psi)$ and $\Omega_F(\Psi)$ if one wants to use~(\ref{eq-EH-bc}) 
to calculate $\Psi_0(\theta)$. We shall return to this important point 
in \S~\ref{subsec-setup}.


\section{Disruption of the hole--disk  magnetic link by the black hole
rotation}
\label{sec-idea}

The main topic of this paper is a force-free magnetic link
between a Kerr black hole and a thin, infinitely conducting
Keplerian accretion disk around it. Thus, we are primarily 
interested in the closed-field configuration depicted 
schematically in Fig.~\ref{fig-geometry-closed}. In contrast 
to the open-field configuration, in which all the field
lines piercing the event horizon extend to infinity, in the 
closed-field configuration, magnetic field lines connect the 
black hole to the disk, forming a nested structure of toroidal 
flux surfaces. In this section we will examine the conditions
under which this configuration can exist and, in particular, 
will discuss the limitations that the rotation of the black hole 
imposes on the radial extent of the force-free magnetic link 
between the disk and the hole.

First, we would like to point out that a magnetically-linked black 
hole--disk system is dramatically different from a magnetically-linked 
star--disk system in at least one important aspect. Indeed, let us 
examine the system's evolution on the shortest relevant, i.e., rotation, 
timescale. In the case where the central object is a highly-conducting 
star, such as a neutron star or a young star, it turns out that no 
steady state configuration with the topology similar to that presented 
in Figure~\ref{fig-geometry-closed} is possible. This is because both 
the disk and the star can be regarded (on this short timescale) as 
perfect conductors, so that the footpoints of the field lines that 
link the two are frozen into their surfaces. Hence, the disk footpoint 
of a given field line rotates with its corresponding Keplerian rotation 
rate, $\Omega_K(r)$, whereas the footpoint of the same field line on the 
star's surface rotates with the stellar angular velocity~$\Omega_*$. 
Therefore, each field line connecting the star to the disk [with the 
exception of a single line connecting to the disk at the corotation 
radius $r_{\rm co}$ where $\Omega(r_{\rm co})=\Omega_*$] is subject 
to a continuous twisting. This twisting results in the generation of 
toroidal magnetic flux out of the poloidal flux, which tends to inflate 
and even open the magnetospheric flux surfaces after only a fraction of 
one differential star--disk rotation period (e.g., van~Ballegooijen~1994;
Uzdensky~et~al. 2002; Uzdensky 2002a,b).

On the other hand, in the case of a black hole being the central object
the situation is different. The key difference is that, unlike stars, 
black holes do not have a conducting surface. On the contrary, they 
are actually effectively quite resistive, in the language of the 
Membrane Paradigm (see Znajek~1978; Damour~1978; MacDonald 
\& Suen 1985; Thorne~et~al. 1986). 
The rather large effective resistivity makes it in principle 
possible for the field lines frozen into a rotating conducting 
disk to slip through the event horizon. This fact makes a quest 
for a stationary closed-field configuration in the black-hole 
case a reasonable scientific task, since it is at least conceivable 
that such configurations may in principle exist.

In our previous paper (Uzdensky~2004) we studied exactly this question 
for the case of a Schwarzschild black hole. We found that a 
stationary force-free configuration of the type depicted in 
Figure~\ref{fig-geometry-closed} indeed exists in this case. 
At the same time, however, there is of course no guarantee 
that a similar configuration will exist in the Kerr case. 
This is because the nonlinear terms in the Grad--Shafranov 
equation that correspond to field-line rotation and toroidal 
field pressure are no longer small in the Kerr case, whereas 
in the Schwarzschild case these terms, although formally finite,
were only at a few per cent level.
In fact, we can make an even stronger statement: even for a 
slowly-rotating Kerr black hole, a force-free configuration 
in which magnetic field connects the polar region of the horizon 
to arbitrarily large distances on the disk (which is precisely
the geometry depicted in Fig.~\ref{fig-geometry-closed}) does 
not exist! We shall now present the basic physical argument
for why this must be the case.

Let us suppose that a force-free configuration of Figure~\ref
{fig-geometry-closed}, where all the field lines attached to 
the disk at all radii thread the event horizon, does indeed exist.
First, let us consider the polar region of the black hole, $r=r_H$, 
$\theta\rightarrow 0$. Suppose that near the rotation axis the flux
$\Psi_0(\theta)$ behaves as a power law: $\Psi_0\sim\theta^\gamma$ 
(the most natural behavior corresponding to a constant poloidal field being 
$\Psi_0\sim\theta^2$). Then note that in a configuration under consideration,
the field lines threading this polar region connect to the disk at some very 
large radius $r_0(\Psi)\gg r_H$. Since the field lines rotate with the 
Keplerian angular velocity of their footpoints on the disk, $\Omega_F(\Psi) 
\sim r_0^{-3/2}(\Psi) \rightarrow 0$ as $\Psi\rightarrow 0$, one finds that, 
for sufficiently small $\Psi$ [and hence sufficiently large $r_0(\Psi)$], 
$\Omega_F(\Psi)$ becomes much smaller than the black hole rotation rate 
$\Omega_H = a/2r_H$. Now let us look at the event horizon regularity 
condition~(\ref{eq-EH-bc}). For the field lines under consideration,
we find that $\sin\theta d\Psi_0/d\theta \sim \theta \theta^{\gamma-1}
\sim \Psi$ and $\delta\Omega = \Omega_F(\Psi) - \Omega_H 
\simeq -\Omega_H = {\rm const} \neq 0$, $\Psi\rightarrow 0$.
Thus,
\beq
I(\Psi) \sim -\Omega_H \Psi \sim -a\Psi,
\qquad {\rm as} \quad \Psi\rightarrow 0 \, ,
\label{eq-I-axis}
\eeq
and, correspondingly,
\beq
II'(\Psi\rightarrow 0) \sim a^2 \Psi \, .
\label{eq-II'-axis}
\eeq

Now, let us look at the force-free balance on the same field lines
but far away from the black hole, at radial distances of the order 
of $r\sim r_0 \gg r_H$. At these large distances $\alpha\approx 1$
and $\delta\Omega \varpi \ll c$, so that the electric terms in the 
Grad--Shafranov equation are small and the coefficient $D$ is close 
to~1. Then the LHS of the Grad--Shafranov equation~(\ref{eq-GS-2}) 
is essentially a linear diffusion-like operator and can be estimated 
as being of the order of $\Psi/r^2$. We see that both the LHS and 
the RHS given by equation~(\ref{eq-GS-2}) scale linearly with $\Psi$ 
but the LHS has an additional factor $\sim r^{-2}$. Thus we conclude
that at sufficiently large distances this term becomes negligible when 
compared with the $II'(\Psi)$-term~(\ref{eq-II'-axis}). In other words, 
the toroidal field, produced in the polar region of the horizon by 
the black hole dragging the field lines along, turns out to be too 
strong to be confined by the poloidal field tension at large distances. 
In fact, this argument suggests that the maximal radial extent $r_{max}$ 
of the region on the disk connected to the polar region of a Kerr black 
hole should scale as $r_{\rm max}\sim r_H/a$ in the limit $a\rightarrow 0$.
One should note that, in the Schwarzschild limit $a\rightarrow 0$, this 
maximal distance goes to infinity and hence a fully-closed force-free 
configuration can exist at arbitrarily large distances, in agreement 
with the conclusions of our paper~I. [Also note that if one tries to 
perform a similar analysis for the Schwarzschild case, then from the 
horizon regularity condition one finds that $I(\Psi)=\Omega_K(\Psi)\sin\theta 
(d\Psi/d\theta)|_{r=r_H} \sim \Psi \cdot r_0^{-3/2}(\Psi)$. Then, assuming 
that $r_0(\Psi)$ is a power law at large distances, $r\sim r_0(\Psi)$, 
the toroidal-field pressure term can be estimated as $II'(\Psi)\sim 
\Psi\cdot r_0^{-3}(\Psi)$. We can thus see that at large distances
this term becomes negligible compared with the LHS ($\sim \Psi r^{-2}$),
so no limitation on the radial extent of the magnetic link can be derived.]

We also would like to remark that this finding is not really surprising 
in view of some important properties axisymmetric force-free magnetospheres, 
known from the general theory of the (non-relativistic) Grad--Shafranov 
equation (see, e.g., van~Ballegooijen~1994; Uzdensky~2002b). This analogy 
is so important that we would like to make a digression to describe it here. 
Let us consider a closed simply-connected (i.e., without magnetic islands) 
axisymmetric configuration like the one shown in Figure~\ref
{fig-geometry-closed}. Then start to increase gradually the overall magnitude 
(which we shall call $\lambda$) of the nonlinear source term $II'(\Psi)$ 
--- the so-called generating function --- starting from zero. As we are 
doing this, let's keep the functional shape of $I(\Psi)$, as well as the 
boundary conditions for $\Psi$, fixed. Then one finds the following 
interesting behavior: there is a certain maximal value $\lambda_{\rm max}$
(whose exact value depends on the details of the functional shape of 
$I(\Psi)$ and the boundary conditions), such that one finds no solutions 
of the Grad--Shafranov equation for $\lambda > \lambda_{\rm max}$. 
For $\lambda<\lambda_{\rm max}$, one actually finds {\it two} solutions  
and these two solutions correspond to two different values of the field-line 
twist angles $\Delta\Phi(\Psi)$. In the limit $\lambda\ll \lambda_{\rm max}$
the two solutions are remarkably different. One of them corresponds to 
$\Delta\Phi\sim\lambda/\lambda_{\rm max} \ll 1$; it is very close to the 
purely potential closed-field configuration and can be obtained as a 
perturbation from the potential solution. The other solution corresponds 
to some finite distribution $\Delta\Phi_c(\Psi)$, in general of order 1 
radian, and is characterized by very strongly inflated poloidal field 
lines. This configuration in fact approaches the open-field geometry
in the limit $\lambda \rightarrow 0$.
Now, as one increases $\lambda$, the difference between the two 
solutions decreases and they in fact merge into one single solution 
at $\lambda=\lambda_{\rm max}$. The corresponding configuration shows 
some modest inflation of the poloidal field and corresponds to the 
field line twist angles that are finite (i.e., of order 1 radian) but 
less than $\Delta\Phi_c(\Psi)$. Most importantly, as we mentioned above, 
no solutions with the required simple topology (i.e., without magnetic 
islands) exist for $\lambda>\lambda_{\rm max}$. 

Clearly, this is exactly what happens in the Kerr black hole case. 
Indeed, in this case the regularity condition~(\ref{eq-EH-bc}) 
requires that the generating function $II'(\Psi)$ be of the order 
of $a^2\Psi$ for small $\Psi$. In a certain sense, the spin parameter 
$a^2$ effectively plays the role of the parameter $\lambda$ from our 
example above. If one considers a configuration in which the magnetic 
link extends to a radius $r_{\rm max}$ on the disk and fixes the disk 
boundary conditions, it turns out that there is a critical maximum value 
$a_{\rm max}^2$ beyond which no solution can be found. From the argument 
presented in the beginning of this section we expect that $a_{\rm max}$ 
scale inversely with $r_{\rm max}$; in particular, for an infinitely 
extended link ($r_{\max}\rightarrow \infty$), one finds $a_{\rm max} 
\rightarrow 0$ and no solution is found for any~$a>0$!

To sum up, even though the field lines can, to a certain degree, 
slip through the horizon because the latter is essentially resistive,
in some situations the horizon is not resistive enough to ensure
the existence of a steady force-free configuration! Indeed, the
field lines are "dragged" by the rotating black hole to such a
degree that, in order for them to slip through the horizon steadily, 
they must have a certain rather large toroidal component. When, for 
fixed disk boundary conditions, the black-hole spin parameter $a$ 
is increased beyond a certain limit $a_{\rm max}(r_{\rm max})$, this 
toroidal field becomes so large that the poloidal field tension is 
no longer able to contain its pressure at large distances.

Finally, the argument put forward in this section proves that 
it may be not only interesting but also in fact necessary to 
consider hybrid configurations in which at least a portion of 
the field lines are open and magnetic disk-hole coupling plays 
a more limited role.


\section{Numerical Simulations}
\label{sec-numerical}

In order to verify the proposition put forward in the preceding
section and to study the magnetically-coupled disk--hole magnetosphere,
we have performed a series of numerical calculations. We obtained 
the solutions of the force-free Grad--Shafranov equation corresponding
to various values of two parameters: the black-hole spin parameter~$a$
and the radial extent $R_s$ of the magnetic link. In this section
we shall describe the actual computational set-up of the problem, 
including the boundary conditions and the numerical procedure; we
shall also present the main results of our calculations.


\subsection{Problem formulation and boundary conditions}
\label{subsec-setup}

We start by describing the basic problem set-up and the boundary
conditions.

The simplest axisymmetric closed-field configuration one could
consider is that shown in Figure~\ref{fig-geometry-closed}. 
In this configuration, all magnetic field lines connect the disk 
and the hole. Furthermore, the entire event horizon and the entire 
disk surface participate in this magnetic linkage; in particular, 
the field lines threading the horizon very close to the axis 
$\theta=0$ are anchored at some very large radial distances 
in the disk:
\beq
\Psi_0(\theta\rightarrow 0) \equiv
\Psi(r=r_H,\theta\rightarrow 0) =
\Psi_{\rm disk}(r\rightarrow \infty) \equiv
\Psi(r\rightarrow \infty, \theta=\pi/2)
\eeq

However, as follows from the arguments presented in \S~\ref{sec-idea},
a steady-state force-free configuration of this type can only exist 
in the case of a Schwarzschild black hole; in the case of a Kerr black
hole, even a slowly-rotating one ($a\ll M$), such a configuration is 
not possible. And indeed, in complete agreement with this point of
view, in our simulations, we were not able to obtain a convergent
solution even for a Kerr black hole with the spin parameter as 
small as $a=0.05$.

Also in \S 3 we proposed a conjecture that, for a given value of $a$,
the magnetic link between the polar region of the black hole and the disk
cannot, generically, extend to distances on the disk larger than a 
certain $r_{\rm max}(a)$. The exact value of $r_{\rm max}$ depends
on the details of the problem, such as the exact flux distribution
$\Psi_d(r)$ on the surface of the disk, etc. However, we proposed
that $r_{\rm max}$ is a monotonically decreasing function of~$a$,
and, more specifically, in the limit $a\rightarrow 0$, $r_{\rm max}$ 
is inversely proportional to~$a$. For a finite ratio $a/M=O(1)$,
we expect that the magnetic link can only be sustained over a 
finite range of radii not much larger than the radius of the 
Innermost Stable Circular Orbit $r_{\rm ISCO}$.

In order to test these propositions, we set up a series of
numerical calculations aimed at solving the Grad--Shafranov 
equation for various values of two parameters: the black-hole 
spin parameter $a$ and the radial extent of the magnetic coupling 
on the disk surface~$R_s$.

Correspondingly, in order to investigate the dependence 
on the radial extent of magnetic coupling, we modified 
the basic geometry of the configuration by allowing for
two topologically-distinct regions: region of closed field 
lines connecting the black hole to the inner part of the disk
$r<R_s$, and the region of open field lines extending from 
the outer part of the disk all the way to infinity.%
\footnote
{In general, open field lines originating from the disk
may carry a magnetocentrifugal wind (Blandford \& Payne 1982)
and the resulting mass-loading may make a full-MHD treatment
necessary for these field lines. Here, however, we shall ignore
this complication and will assume the force-free approach to be
valid in this part of the magnetosphere as well.}
This configuration is shown in Figure~\ref{fig-geometry-kerr}. 
We count the poloidal flux on the disk from the radial 
infinity inward, so that $\Psi_d(r=\infty)=0$, and 
$\Psi_d(r=r_H)=\Psi_{\rm tot}$. The disk flux distribution 
may still be the same as in the configuration of 
Figure~\ref{fig-geometry-closed}; however, now there is a critical
field line $\Psi_s\equiv \Psi_d(R_s)<\Psi_{\rm tot}$ that
acts as a separatrix between open field lines ($\Psi<\Psi_s$) 
and closed field lines ($\Psi_s<\Psi<\Psi_{\rm tot}$)
connecting to the black hole. Correspondingly, the poloidal
flux on the black hole surface varies from $\Psi=\Psi_s$
at the pole $\theta=0$ to $\Psi=\Psi_{\rm tot}$ at the
equator $\theta=\pi/2$.

It is worth noting that a more general configuration
would also have some open field lines connecting the
polar region of the black hole to infinity. In fact,
such a configuration would be more physically interesting 
because these open field lines would enable an additional
extraction of the black hole's rotational energy via the 
Blandford--Znajek mechanism (BZ77). We shall call this 
a hybrid configuration because the disk--hole magnetic
coupling and the Blandford--Znajek mechanism operate
simultaneously. In the present paper, however, we shall
assume that no such hole--infinity open field lines. 
We make this choice not because of any physical reasons
but simply because of technical convenience: we want to 
isolate the effect tof disk-hole coupling. In addition, 
as we discuss in more detail in \S~\ref{sec-conclusions}, 
a proper treatment of these open field lines would require
a more complicated numerical procedure than that needed 
for the field lines that connect to the conducting disk.

In addition to boundary conditions, one has to specify the 
angular velocity $\Omega_F(\Psi)$ of the magnetic field lines. 
Since we assume that the disk is a perfect conductor, and since
in our field configuration all the field lines go through the 
disk, this angular velocity is equal to that of the matter in
the disk.

Now let us consider the open field lines $\Psi< \Psi_s$.
In principle, since they are attached to a rotating Keplerian 
disk, these lines rotate differentially with the angular velocity
$\Omega_F(\Psi)=\Omega_K[r_0(\Psi)]$. Correspondingly, just as
the closed field lines going into the black hole or the open
field lines in a pulsar magnetosphere, they have to cross a 
light cylinder 
and therefore have to carry poloidal current $I(\Psi)$, whose
value must be consistent with, and indeed determined by, the 
regularity condition at the light cylinder. Because this outer
light cylinder is very distinct from the inner light cylinder 
that is crossed by the closed field lines entering the event horizon, 
we in general would expect the function $I(\Psi<\Psi_s)$ be very
different from the function $I(\Psi>\Psi_s)$. In particular, we
would expect a discontinuous behavior, $I_s^{\rm open}\equiv 
\lim\limits_{\Psi\rightarrow\Psi_s} I(\Psi<\Psi_s) \neq
I_s^{\rm closed}\equiv \lim\limits_{\Psi\rightarrow\Psi_s} I(\Psi>\Psi_s)$,
even though the field-line angular velocity $\Omega_F=\Omega_K[r_0(\Psi)]$
remains perfectly continuous and smooth at $\Psi=\Psi_s$.

Dealing with such a discontinuity in $I(\Psi)$ across the separatrix 
$\Psi=\Psi_s$ presents certain numerical difficulties, especially
taking into account that the location of the separatrix $r_s(\theta)=
r(\Psi=\Psi_s,\theta)$ is not known a priori. Therefore, in the present
study we decided to simplify the problem by introducing the following
modifications: we require that the outer part of the disk, $r>R_s$,
be nonrotating: $\Omega_F(\Psi<\Psi_s)\equiv 0$.

Correspondingly, the open field lines do not cross an outer light 
cylinder, and so $I(\Psi<\Psi_s)\equiv 0$. To put it in other words, 
we just take the open-field outer part of the disk magnetosphere
to be potential. Next, in order to avoid the numerically-challenging
discontinuities in $\Omega_F(\Psi)$ and $I(\Psi)$ at $\Psi=\Psi_s$,
we slightly modify the disk rotation law just inside of $R_s$ 
by taking $\Omega_F$ smoothly to zero over a small (compared 
with the total amount of closed flux) poloidal flux range. In 
particular, we used the following prescription:
\begin{eqnarray}
\Omega_F(\Psi) &=& 0\, , \qquad \Psi<\Psi_s \, , \nonumber \\
\Omega_F(\Psi) &=& \Omega_K[r_0(\Psi)] \cdot
\tan^2 \bigl({{\Psi-\Psi_s}\over{\Delta\Psi}}\bigr)\, , \qquad 
\Psi>\Psi_s \, ,
\label{eq-OmegaofPsi}
\end{eqnarray}
where $\Delta\Psi=0.2(\Psi_{\rm tot}-\Psi_s)$ and
(see equation [5.72] of Krolik~1999, p.~117)
\begin{equation}
\Omega_K(r) = {\sqrt{M}\over{r^{3/2}+a\sqrt{M}}} \, .
\label{eq-Keplerian}
\end{equation}

These modifications enabled us to focus on examining how 
black hole rotation (i.e., the spin parameter $a$) limits 
the radial extent $R_s$ of the force-free magnetic coupling,
while at the same time avoiding certain numerical difficulties
resulting from the discontinuous behavior of poloidal current
$I(\Psi)$. We believe that these modifications do not lead to
any significant qualitative change in our conclusions, especially
in the case of small $a$ and large $R_s$. Nevertheless, we intend
in the future to enhance our numerical procedure so that it become
fully capable of treating this discontinuity.

Let us now describe the computaional domain and the boundary conditions.

First, because of the assumed axial symmetry and the symmetry 
with respect to the equatorial plane, we performed our computations
only in one quadrant, described by $\theta\in[0,\pi/2]$ and
$r\in[r_H,\infty]$. Thus, we have four natural boundaries of the domain:
the axis $\theta=0$, the infinity $r=\infty$, the equator 
$\theta=\pi/2$, and the horizon $r=r_H$. Of these, the axis 
and the equator require boundary conditions for $\Psi$, whereas
the horizon and the infinity are actually regular singular surfaces
and so we only impose regularity conditions on them.

The boundary condition on the rotation axis is particularly simple:
\beq
\Psi(r,\theta=0) = \Psi_s = {\rm const} \, .
\label{eq-bc-axis}
\eeq

The equatorial boundary, $\theta=\pi/2$, actually consists
of two parts: the disk (considered to be infinitesimally thin) 
and the plunging region between the disk and the black hole.
The border between them, i.e., the inner edge of the disk, is
assumed to be very sharp and to lie at the ISCO: 
$r_{\rm in}=r_{\rm ISCO}(a)$; $r_{\rm ISCO}$ varies between 
$r_{\rm ISCO}=6M$ for a Schwarzschild black hole ($a=0$)
and $r_{\rm ISCO}=M$ for a maximally rotating Kerr black hole
($a\rightarrow 1$).

Let us first discuss the boundary conditions at the disk surface,
$r>r_{\rm in}$. Depending on the resistive properties of the disk,
and on the timescale under consideration, one can choose between 
two possibilities, both of which appear to be physically sensible: 

1) If one is interested in time-scales much longer than the
characteristic rotation timescale but much shorter than both 
the accretion and the magnetic diffusion timescales, then it
is reasonable to regard the poloidal flux distribution across 
the disk to be a fixed prescribed function, which must be
specified explicitly. Thus, in this case one adopts a Dirichlet-type
disk boundary condition:
\beq
\Psi(r>r_{\rm in},\theta=\pi/2) = \Psi_d(r)
\label{eq-bc-disk-Dirichlet}
\eeq
The function $\Psi_d(r)$ is arbitrary; the only requirement that
must be imposed in accordance with the discussion above is the
convention that $\Psi_d(r=\infty)=0$ and $\Psi_d(r_{\rm in})=
\Psi_{\rm tot}$. Since we don't have any good physical reasons
to favor one choice of $\Psi_d(r)$ over any other, we in this 
paper just choose it arbitrarily to be a power-law with the
exponent equal to $-1$:
\beq 
\Psi_d(r) = \Psi_{\rm tot}\, \biggl({{r_{\rm in}}\over r}\biggr) \, .
\label{eq-Psi_d}
\eeq

2) If one looks for a configuration that is stationary on timescales
much longer than the effective magnetic diffusion time (while perhaps
still much shorter than the accretion time scale), then one should
regard the disk as effectively very resistive for the purposes of
specifying the disk boundary condition. This situation may arise 
in the case of a turbulent disk; for such a disk, the effective
magnetic diffusivity $\eta$ can probably be estimated as 
$\eta_{\rm turb}=\alpha_{SS} c_s h$, in the spirit of the 
$\alpha$-prescription for the effective viscosity in the 
SS73 model. Then, the characteristic radial velocity of 
the magnetic footpoints across the disk is roughly 
$v_{\rm fp} \sim \alpha_{SS} c_s (B_r/B_z)_d$. For the
ratio $(B_r/B_z)_d$ of order 1, this velocity is much 
greater (by a factor of $r/h$) than the characteristic 
accretion velocity. Therefore, the only way one can have 
a steady-state configuration on the diffusion time-scale
(which, according to the above estimate is of the order of the
disk sound crossing time $r/c_s$) is for the poloidal field
to be nearly perpendicular to the disk, $B_r \ll B_z$. This
requirement translates into a simple von-Neumann boundary
condition for $\Psi(r,\theta)$ at the disk surface:
\beq
{{\partial\Psi}\over{\partial\theta}} (r,\theta={\pi\over 2}) = 0 \, .
\label{eq-bc-disk-Neumann}
\eeq
In our present paper, however, we chose the Dirichlet-type boundary condition 
represented by equations~(\ref{eq-bc-disk-Dirichlet})--(\ref{eq-Psi_d}) and
set $\Psi_{\rm tot}=1$ throughout the paper.

In the plunging region $(r_H\leq r\leq r_{\rm in},\theta=\pi/2)$
we have chose 
\beq
\Psi(r_H\leq r\leq r_{\rm in},\theta=\pi/2)=\Psi_{\rm tot} \equiv
\Psi_d(r_{\rm in}) = {\rm const} \, .
\label{eq-bc-plunging}
\eeq
This choice appears to be physically appropriate for an accreting 
(and not just rotating) disk. The reason for this is that the 
matter in this region falls rapidly onto the black hole and 
thereby stretches the magnetic loops in the radial as well 
as the azimuthal directions, greatly reducing the strength 
of the vertical field component. The horizontal magnetic 
field then reverses across the plunging region, which is 
thus described as an infinitesimally thin non-force-free
current sheet lying along the equator. In essense, this 
situation is directly analogous to the case of a force-free 
pulsar magnetosphere, where all the field lines crossing the 
outer light cylinder have to be open and extend out to infinity, 
thus forming an equtorial current sheet (Beskin~2003; van~Putten 
\& Levinson 2003). 
In the black-hole case, one could still consider an alternative 
picture of the plunging region with some field lines crossing 
the equator inside $r_{\rm in}$. However, in this case one would 
still have to have a non-FFDE equatorial current sheet inside 
the inner light cylinder, as was shown by Komissarov (2002b, 2004a).

Finally, as we have discussed in~\S~\ref{subsec-EH}, the event 
horizon is a regular singular surface of the Grad--Shafranov equation.
Correspondingly, one cannot and need not impose an additional arbitrary 
boundary condition here (e.g., Beskin \& Kuznetsova~2000; Komissarov~2002b, 
2004a). Instead, one imposes the regularity condition~(\ref{eq-EH-bc}); 
this condition has the form of an ordinary differential equation (ODE) 
that determines the function $\Psi_0(\theta)$ provided that both 
$\Omega_F(\Psi)$ and $I(\Psi)$ are given. Thus, from the procedural 
point of view, this condition can be used as a Dirichlet boundary 
condition on the horizon. It is important to acknowledge, however, 
that one does not have the freedom of specifying an arbitrary 
function $\Psi(\theta)$ and then studying how the information 
contained in this function propagates outward and affects the 
solution away from the horizon. The function $\Psi_0(\theta)$ 
is uniquely determined once $\Omega_F(\Psi)$ and $I(\Psi)$ are 
given and thus there is no causality violation here. 

Similarly, the spatial infinity $r=\infty$ is also a regular singular 
surface of the Grad--Shafranov equation and thus can also be described 
by a regularity condition. In this sense, the horizon and the infinity 
are equivalent (e.g., Punsly \& Coroniti 1990). Note that, in our 
particular problem set-up, the sitiuation at infinity is greatly 
simplified because we have set $\Omega_F(\Psi)=0$ on the open field 
lines extending from the disk. Because of this, there is no outer 
light cylinder for these lines to cross and thus one can also set 
$I(\Psi<\Psi_s)=0$. Then, at very large distances ($r\gg r_H$), 
the Grad--Shafranov equation~(\ref{eq-GS-2}) becomes a very simple 
linear equation:
$\Psi_{rr}+r^{-2}\sin\theta \partial_\theta(\Psi_\theta/\sin\theta) = 0$,
and the asymptotic solution that corresponds to the open field geometry
with finite magnetic flux is just 
\beq
\Psi(r=\infty,\theta)=\Psi_s \cos\theta \, .
\label{eq-bc-infty}
\eeq


\subsection{Light-Cylinder Regularity Condition}
\label{subsec-LC}

At this point the problem is almost completely determined.
The only thing we still have to specify is the poloidal
current $I(\Psi)$. Unlike $\Omega_F(\Psi)$, which was determined
from the frozen-in condition on the disk surface, the function
$I(\Psi)$ cannot be explicitly prescribed as an arbitrary
function on any given surface. Instead, it must be somehow
determined self-consistently together with the solution $\Psi(r,\theta)$ 
itself. This means that there must be one more condition that 
we have not yet used. And indeed, this  
additional condition is readily found --- it is the (inner)
light-cylinder regularity condition. Let us look at it more closely.

As can be easily seen from the Grad--Shafranov equation~(\ref{eq-GS-2}), 
the light cylinder, defined as the surface where 
\beq
D=0 \quad \Rightarrow \quad \alpha=\alpha_{\rm LC} =
|\delta\Omega| \varpi\, , 
\label{eq-LC}
\eeq
is a singular surface, because the coefficients in front of 
both the $r$- and $\theta$- second-order derivatives of~$\Psi$
vanish there. Physically speaking, the light cylinder is the 
surface where the locally-measured rotational velocity of the 
magnetic field lines with respect to the ZAMOs is equal to the 
speed of light, $v_{\rm B,\phi}=c$, and where $E=B_{\rm pol}$ 
in the ZAMO frame. In general relativity there are two light 
cylinders, the inner one and the outer one. The outer light 
cylinder is just a direct analog of the pulsar light cylinder; 
it is crossed by rotating field lines that are open and extend 
to infinity. In our problem, we are interested in the closed field 
lines, i.e., those reaching the event horizon. These field lines 
cross the so-called inner light cylinder, whose existence is a
purely general-relativistic effect, first noticed by Znajek (1977)
and by~BZ77.

Because the inner light cylinder is a singular surface of 
equation~(\ref{eq-GS-2}), in general this equation admits 
solutions that are not continuous or continuously differentiable 
at the light cylinder. Such solutions, while admissible mathematically,
are not physically possible. Thus, we supplement our mathematical problem 
by an additional physical requirement that the solution be continuous and 
smooth across the light cylinder surface. In particular, this means that 
the 1st and 2nd derivatives of~$\Psi$ must be finite there. Correspondingly,
one can just drop all the terms proportional to~$D$ when applying equation
(\ref{eq-GS-2}) at the light cylinder and keep only the terms involving 
the derivatives of~$D$. The result can be formulated as an expression 
that determines the function $I(\Psi)$, namely:
\beq
-II'(\Psi)={\Delta\over{\rho^2}} \Psi_r (\partial_r D)|_{\rm LC}+
{1\over{\rho^2}} \Psi_\theta (\partial_\theta D)|_{\rm LC} +
\delta\Omega \Omega_F' \varpi^2 (\nabla\Psi)^2 |_{\rm LC}\, ,
\label{eq-LC-regularity}
\eeq
where $\Psi$, $r$, and $\theta$ are taken at the light cylinder:
\beq
\Psi=\Psi_{\rm LC}(\theta)=\Psi[r_{\rm LC}(\theta),\theta] \, ,
\label{eq-Psi_LC}
\eeq
and the function $r_{\rm LC}(\theta)$ ---the shape of the light 
cylinder surface --- is determined implicitly by equation~(\ref{eq-LC}).
This approach was first used successfully at the outer light cylinder
by Contopoulos~et~al. (1999) in the context of pulsar magnetspheres.
In the black hole problem, it was first used by Uzdensky (2004) for
the Schwarzschild case.

Let us now discuss how one can use the light cylinder regularity 
condition~(\ref{eq-LC-regularity}) to determine $I(\Psi)$ in practice. 
Conceptually, one can think of this condition as follows. Suppose one 
starts by fixing all the other boundary and regularity conditions in 
the problem [including the choice of $\Omega_F(\Psi)$]. Then, for an 
arbitrarily chosen function~$I(\Psi)$, one can regard the condition~(\ref
{eq-LC-regularity}) as a mixed-type, Dirichlet-Neumann boundary condition
because it can be viewed as a quadratic algebraic equation for, say, the 
first radial derivative. Thus, if $I(\Psi)$ is given, one can express 
$\Psi_r|_{\rm LC}$ in terms of $\Psi_{\rm LC}$ and $\Psi_\theta|_{\rm LC}$. 
Next, one applies this condition separately on each side of the light 
cylinder and gets a complete, well-defined problem in each of the two 
regions separated by the light cylinder. Then, one can obtain a solution 
in each of these regions. Because of the use of the regularity 
condition~(\ref{eq-LC-regularity}), each of the two solutions 
is going to be regular near the light cylinder. In general, 
however, these solutions are not going to match each other 
at $r=r_{\rm LC}(\theta)$ and the mismatch $\Delta\Psi_{\rm LC}(\theta)$ 
will depend on the original choice of the function~$I(\Psi)$. 
This observation suggests a method for selecting a unique 
function~$I(\Psi)$: one can devise a procedure in which one 
iterates with respect to $I(\Psi)$ until $\Delta\Psi_{\rm LC}$ 
becomes zero. The corresponding function $I(\Psi)$ is then declared 
the correct one: only with this choice of $I(\Psi)$ the solution 
$\Psi(r,\theta)$ passes smoothly through the light cylinder. 

The above method for determining $I(\Psi)$ is conceptually 
illuminating and can be easily implemented in simple cases.
For example, in the case of a {\it uniformly-rotating} pulsar
magnetosphere, two important simplifications take place. First,
the location of the light cylinder is known a priori, $r_{\rm LC}(\theta)=
c/\Omega=\rm const$, and hence one can choose a computational grid that 
is most suitable for dealing with the light cylinder  (e.g., cylindrical 
polar coordinates with some gridpoints lying on the cylinder).
Second, because $\Omega_F=\rm const$, the terms quadratic in 
the derivatives of $\Psi$ disappear, and the task of resolving
equation (\ref{eq-LC-regularity}) with respect to the derivative 
normal to the light cylinder becomes trivial. These simplifications 
make the procedure described above very practical and it was in fact 
used successfully by Contopoulos~et~al. (1999) (and repeated later by 
Ogura \& Kojima 2003) to obtain a unique solution for an axisymmetric 
pulsar magnetosphere that was smooth across the outer light cylinder.

In the problem considered in this paper, however, the situation is
much more complicated. In particular, the light cylinder's position
and shape are not known a priori; instead, they need to be determined 
self-consistently as part of the solution. Also, equation~(\ref
{eq-LC-regularity}) is, in general, quadratic with respect to 
$\partial_r\Psi$, and hence one has to deal with the problem of 
the existence of its solutions and with the task of selecting only 
one of them. Because of this overall complexity, we decided against 
using this procedure in our calculations. Instead, we chose a much 
simpler and more straight-forward method: we used equation~(\ref
{eq-LC-regularity}) to determine~$I(\Psi)$ [or, rather, the 
combination $II'(\Psi)$ that is actually needed for further 
computations] directly, by explicitly interpolating all the 
terms on the right-hand-side of equation~(\ref{eq-LC-regularity}).
We will describe this in more detail in the next section.


\subsection{Numerical procedure}
\label{subsec-procedure}

We performed our calculations in the domain $\{r\in[r_H,\infty],
\theta\in[0,\pi/2]\}$ on a grid that was uniform in $\theta$ and 
in the variable $x\equiv \sqrt{r_H/r}$ (which enabled us to extend 
the computational domain to infinity). The highest resolution used 
was 60 gridzones in the $\theta$-direction and 200 gridzones in 
the radial ($x$) direction. To solve the elliptic Grad--Shafranov 
equation~(\ref{eq-GS-2}), we employed a relaxation procedure similar
to the one employed by Uzdensky~et~al. (2002). In this procedure, we 
introduced artificial time variable~$t$ and evolved the flux function 
according to the parabolic equation
\beq
{{\partial\Psi}\over{\partial t}} =
\pm f(r,\theta) \bigl(LHS-RHS\bigr)\, ,
\label{eq-relaxation} 
\eeq
where $LHS$ and $RHS$ are the left- and right-hand sides of the 
Grad--Shafranov equation~(\ref{eq-GS-2}), respectively, and the 
factor $f(r,\theta)$ was an artificial multiplier introduced in
order to accelerate convergence in regions where the diffusion 
coefficients in the $x$ and~$\theta$ directions are small (e.g., 
very far away or very close to the horizon). The sign in front 
of $f(r,\theta)$ was chosen according to the sign of the diffusion
coefficient in equation~(\ref{eq-GS-2}): it was plus outside the 
light cylinder (where $D>0$) and minus inside (where $D<0$). 
It is clear that any steady-state configuration achieved as 
a result of this evolution is a solution of the Grad-Shafranov 
equation~(\ref{eq-GS-2}).

Here we would like to draw attention to the following non-trivial problem. 
During the relaxational evolution described by equation~(\ref{eq-relaxation}), 
the light cylinder generally moves across the grid and, from time to time
inevitably gets close to some of the gridpoints. This leads to the danger, 
first noted by Macdonald~(1984), that some gridpoints will oscillate 
between the two sides of the light cylinder. Indeed, suppose that a 
given gridpoint~$P$ is initially on the outer side of the light 
cylinder ($D_P>0$), and so $\partial\Psi_P/\partial t$ is determined 
by equation~(\ref{eq-relaxation}) with the plus sign. Let us suppose 
that the resulting evolution of $\Psi_P$ is such that $D_P$ decreases. 
The, after some time one may find that $D_P$ has become negative; 
correspondingly, at the next timestep one uses equation~(\ref{eq-relaxation}) 
with the minus sign and so $\Psi_P$ starts to evolve in the opposite direction.
Because the value of $D$ at a fixed spatial point~$P$ is, locally, a smooth 
monotonic function of~$\Psi_P$, it now starts to increase and may become 
positive again in one or two timesteps. This leads to rapid small-amplitude
oscillations of the light cylinder around some gridpoints, instead of a
smooth large-scale motion associated with the iteration process. As a 
result, the light cylinder gets "stuck" on these gridpoints and the 
function $r_{\rm LC}(\theta)$ becomes a series of steps and plateaus
instead of a smooth curve. A simple and efficient way to avoid this 
problem turned out to be to update the function $D(r,\theta)$ not at 
every timestep but rather very infrequently. Although it caused some 
delay in the convergence of the relaxation process, this modification 
has worked very well in practice, enabling the light cylinder to move 
freely across the grid and to achieve its ultimate smooth shape. 

To implement our relaxation procedure numerically, we used an 
explicit finite-difference scheme with 1st-order accurate time 
derivative and centered 2nd-order accurate spatial derivatives. 
It is also worth mentioning that writing out the full-derivative 
terms such as $\partial_r (\Psi_r D\Delta/\rho^2)$ as $\Psi_r 
\partial_r (D\Delta/\rho^2)+ (D\Delta/\rho^2)\Psi_{rr}$, etc., 
and then evaluating them on the grid actually worked better 
than evaluating these full derivatives directly as they are.
The initial condition---the starting point of our relaxation 
process---was prescribed explicitly as 
\beq
\Psi(t=0,r,\theta) = \Psi_s + [\Psi_d(r)-\Psi_s] \, (1-\cos\theta) \, .
\label{eq-initial-condition}
\eeq
Also, we found it useful to use cubic-spline interpolation 
of functions $I(\Psi)$ and~$\Omega_F(\Psi)$ to avoid some 
small-scale rapid oscillations of the solution.

Finally, let us describe the particular numerical implementation of 
the procedure that was used to determine the poloidal current~$I(\Psi)$ 
in our code. As we mentioned at the end of the previous section, we used 
equation~(\ref{eq-LC-regularity}) explicitly to determine $II'(\Psi)$ by 
interpolating all the terms on the right-hand side of that equation at 
the light cylinder. Because the light cylinder surface is roughly spherical, 
it was convenient to represent $I(\Psi)$ by a tabular function specified on 
a one-dimensional array $\{\Psi_{\rm LC}^j\}$ of the values of $\Psi_{\rm LC}$ 
at the radial rays $\theta=\theta^j= jh_\theta$, where $h_\theta$
is the grid-spacing in $\theta$. Along each of these rays,
we first had to locate the pair of radially-adjacent gridpoints
between which the light cylinder lay. Then we used an interpolation 
of $D(r,\theta)$ to determine the position $r=r_{\rm LC}(\theta^j)$
of the light cylinder more precisely and to obtain $\Psi_{\rm LC}^j=
\Psi_{\rm LC}(\theta^j)$, as well as the values of the derivatives 
$\Psi_r$, $\Psi_\theta$, $D_r$, and~$D_\theta$ at the light cylinder
for each of the rays. Finally, we used (\ref{eq-LC-regularity}) to 
compute the value of $II'(\Psi)$ at each $\Psi_{\rm LC}^j$. 
This is actually not as trivial as it may seem, because the 
condition~(\ref{eq-LC-regularity}) in such an approach was 
enforced at all times during the relaxation procedure that 
determined $\Psi(r,\theta)$, whereas the Grad--Shafranov equation 
itself was satisfied only after convergence had been reached. 
Therefore, one had to exercise extra care, for example, in 
deciding how often $I(\Psi)$ needs to be updated. We found 
that it was necessary to update $I(\Psi)$ only fairly 
infrequently during our relaxation procedure.


\subsection{Results}
\label{subsec-results}

The single most important result of the present study is 
presented in Figure~\ref{fig-a_of_Psi_s}. This figure shows 
where in the two-dimensional $(a,\Psi_s)$ parameter space 
force-free solutions exist and where they do not. Filled 
circles on this plot represent the runs in which convergence 
was achieved (allowed region), whereas open circles correspond 
to the runs that failed to converge to a suitable solution 
(forbidden region). The boundary $a_{\rm max}(\Psi_s)$ between 
the allowed and forbidden regions is located somewhere inside the
narrow hatched band that runs from the lower left to the upper 
right of the Figure [the finite width of the band represents the
uncertainty in $a_{\rm max}(\Psi_s)$ due to a limited number of 
runs]. As we can see, $a_{\rm max}(\Psi_s)$ is a monotonically 
increasing function. In particular, in the limit $\Psi_s \rightarrow 0$, 
$a_{\rm max}$ indeed scales linearly with $\Psi_s$ and hence is
inversely proportional to $R_s=r_{\rm in} \Psi_{\rm tot}/\Psi_s$, 
in full agreement with our expectations presented in \S~\ref{sec-idea}. 
However, this linear dependence no longer holds for finite values 
of~$\Psi_s$ (and hence of~$a_{\rm max}$).

In order to study the effect that black hole spin has on the solutions, 
we concentrate on several values of~$a$ for a fixed value of~$\Psi_s$. 
In particular, we choose $\Psi_s=0.5$ [that corresponds to $R_s=
2r_{\rm in}(a)$] and considered four values of`$a$: $a=0$, $a=0.25$, 
$a=0.5$, and $a=0.7$. Thus, Figure~\ref{fig-contour} shows the contour 
plots of the poloidal magnetic flux for these four cases. We see that 
the flux surfaces inflate somewhat with increased~$a$, but this expansion 
is not very dramatic, even in the case $a=0.7$, which is very close to 
the critical value $a_{\rm max}(\Psi_s=0.5)$ that corresponds to a sudden 
loss of equilibrium. We note that this finding is completely in line with 
our discussion in \S~\ref{sec-idea}.

The next three Figures present the plots of three important 
functions that characterize the solutions. In each Figure
there are four curves corresponding to our selected values 
$a=0$, 0.25, 0.5, and 0.7 for $\Psi_s=0.5$. Figure~\ref{fig-Psi_0} 
shows the event-horizon flux distribution $\Psi_0(\theta)$;
Figure~\ref{fig-I} shows the poloidal-current function $I(\Psi)$;
and Figure~\ref{fig-alpha_LC} shows the position of the inner light 
cylinder described in terms of the lapse function $\alpha_{\rm LC}(\theta)$.

In Figure~\ref{fig-Psi_0} we also plot $\Psi_0^{(0)}(\theta)$ 
corresponding to the simple split-monopole solution with uniform
radial field at the horizon. Note that on the horizon we have 
$\Delta=0$ and hence $\Sigma=r_H^2+a^2 = 2Mr_H$; therefore 
\beq
B_{\hat{r}}={1\over{\varpi\rho}} \Psi_\theta = 
{1\over{\Sigma\sin\theta}} \Psi_\theta =
{1\over{2Mr_H}}\, {1\over{\sin\theta}}\, \Psi_\theta \, ,\qquad r=r_H\, .
\eeq
Thus, $B_{\hat{r}}(\theta)={\rm const}$ corresponds to 
$\Psi_0^{(0)}(\theta)=\Psi_s+(\Psi_{\rm tot}-\Psi_s) (1-\cos\theta)$,
independent of~$a$. This function is plotted in Figure~\ref{fig-Psi_0}
(the dashed line) for comparison with the actual solutions.
We see that the deviation from $\Psi_0^{(0)}(\theta)$ becomes 
noticeable only when $a$ approaches~1.

Figure~\ref{fig-alpha_LC} shows $\alpha_{\rm LC}(\theta)$. 
An interesting feature here is that the light cylinder reaches 
the event horizon at some intermediate angle $0<\theta_{\rm co}<\pi/2$ 
for small values of~$a$. This is because, when $a<0.359..\, M$, 
the inner edge of a Keplerian disk rotates faster than the black hole; 
correspondingly, somewhere in the disk there exists a corotation point 
$r_{\rm co}> r_{\rm in}$ such that $\Omega_K(r_{\rm co})=\Omega_H$. 
The field line $\Psi_{\rm co}$ threading the disk at this point 
corotates with the black hole. Therefore, at the point $\theta_{\rm co}$ 
where this line intersects the horizon, we have $\delta\Omega=0$, and so 
this point ($r=r_H,\theta=\theta_{\rm co}$) has to lie on the light 
cylinder. The location $\theta_{\rm co}$ of this point moves towards 
the equator when $a$ is increased and reaches it at $a=0.359..\, M$. 
For larger values of $a$, the entire disk outside of the ISCO rotates 
slower than the black hole and the light cylinder touches the horizon 
only at the pole $\theta=0$.

Finally, we also computed all the electric and magnetic field components 
and checked that $E^2<B^2$ everywhere outside the horizon.


\section{Astrophysical Implications/Consequences}
\label{sec-implications}

In this section we'll discuss the exchange of energy and 
angular momentum between the black hole and the disk. Apart
from the question of existence of solutions, this issue is
one of the most important for actual astrophysical applications.
Fortunately, once a particular solution describing the 
magnetosphere is obtained, computing the energy and angular 
momentum transported by the magnetic field becomes very simple.

Indeed, according to MT82, angular momentum and red-shifted
energy (i.e., "energy at infinity") are transported along the 
poloidal field lines through the force-free magnetosphere 
without losses. Thus, the amount of angular momentum $\Delta L$
transported out in a unit of global time $t$ through a region 
between two neighboring poloidal flux surfaces, $\Psi$ and 
$\Psi+\Delta\Psi$, as given by equation~(7.6) of MT82 (modified 
to suit our choice of notation), is
\begin{equation}
{{d\Delta L}\over{dt}} = - {1\over 2}\, I \Delta\Psi \, ,
\label{eq-torque-MT82}
\end{equation}
and the red-shifted power---flux of redshifted energy per unit
global time~$t$--- is expressed as
\begin{equation}
\Delta P =  - {1\over 2}\, \Omega_F I \,  \Delta\Psi \, ,
\label{eq-power-MT82}
\end{equation}
(see eq. [7.8] of MT82).

Then, taking into account the contributions from both hemispheres
and both sides of the disk, we can compute the total magnetic torque 
exerted by the hole onto the disk per unit $t$ as 
\begin{equation}
{{dL}\over{dt}} = -\int\limits_{\Psi_s}^{\Psi_{\rm tot}} I(\Psi) d\Psi \, ,
\label{eq-torque-total}
\end{equation}
and, correspondingly, the total red-shifted power transferred from the 
hole onto the disk via Poynting flux is
\begin{equation}
P = -\int\limits_{\Psi_s}^{\Psi_{\rm tot}} \Omega_F(\Psi) I(\Psi) d\Psi \, .
\label{eq-power-total}
\end{equation}

Next, since in our problem we have an explicit mapping~(\ref{eq-Psi_d}) 
between $\Psi$ and the radial coordinate $r$ on the disk, we can 
immediately write down expressions for the radial distributions
of angular momentum and red-shifted energy deposited on the disk 
per unit global time:
\begin{equation}
{{d\Delta L(r)}\over{dt}} =
- I[\Psi_d(r)]\, {{d\Psi_d}\over dr} \, dr \, ,
\label{eq-torque-of-r}
\end{equation}
and
\begin{equation}
\Delta P(r) = -\Omega_K(r)\, I[\Psi_d(r)]\, {{d\Psi_d}\over dr} \, dr \, .
\label{eq-power-of-r}
\end{equation}

Figures \ref{fig-L-of-r} and~\ref{fig-P-of-r} show these distributions 
for our selected cases $a=0.25$, 0.5, and~0.7 for fixed $\Psi_s=0.5$. 
We see that in the case $a=0.25$ there is a corotation point $r_{\rm co}$
on the disk such that $\Omega_{\rm disk}>\Omega_H$ inside $r_{\rm co}$ 
and $\Omega_{\rm disk}<\Omega_H$ outside $r_{\rm co}$. Correspondingly,
both angular memontum and red-shifted energy flow from the inner
($r<r_{\rm co}$) part of the disk to the black hole and from the 
hole to the outer ($r>r_{\rm co}$) part of the disk. At larger values 
of~$a$, however, the Keplerian angular velocity at $r=r_{\rm in}$ is 
smaller than the black hole's rotation rate and there is no corotation 
point; correspondingly, both angular momentum and redshifted energy 
flow from the hole to the disk. Also, as can be seen in Figures~\ref
{fig-L-of-r} and~\ref{fig-P-of-r}, the deposition of these quantities 
becomes strongly concentrated near the disk's edge, especially at higher
values of~$a$.

Next, Figures~\ref{fig-L-of-a} and~\ref{fig-P-of-a} demonstrate the 
dependence of the total integrated angular momentum and red-shifted 
energy fluxes~(\ref{eq-torque-total})--(\ref{eq-power-total}) on the 
black hole spin~$a$ for a fixed value of $\Psi_s=0.5$. We see that 
both quantities are negative at small values of~$a$ (meaning a transfer
from the disk to the hole,) but then increase and become positive
at larger~$a$. The angular momentum transfer rate depends roughly 
linearly on~$a$, whereas the red-shifted power~$P(a)$ grows even 
faster, especially at large values of~$a$. It is also interesting 
to note that the two quantities go through zero at slightly different
values of the spin parameter: $dL/dt$ becomes zero at $a\approx 0.23$, 
while $P=0$ at $a\approx 0.26$. This means that it possible to have
the total angular momentum flow from the hole to the disk and the 
total power flow from the disk to the hole at the same time.


\section{Conclusions}
\label{sec-conclusions}

In this paper we investigated the structure and the conditions 
for the existence of a force-free magnetosphere linking a rotating 
Kerr black hole to its accretion disk. We assumed that the 
magnetosphere is stationary, axisymmetric, and degenerate and that
the disk is thin, ideally conducting, and Keplerian and that it is 
truncated at the Innermost Stable Circular Orbit. Our main goal was 
to determine under which conditions a force-free magnetic field can 
connect the hole directly to the disk and how the black hole rotation 
limits the radial extent of such a link on the disk surface. 

We first introduced (in \S~\ref{sec-idea}) a very simple but robust 
physical argument that shows that, generally, magnetic field lines 
connecting the polar region of a spinning black hole to arbitrarily 
remote regions of the disk cannot be in a force-free equilibrium.
The basic reason for this can be described as follows. Since the 
field lines threading the horizon have to first cross the inner 
light cylinder, and since they generally rotate at a rate that is 
different from the rotation rate of the black hole, these field 
lines have to be bent somewhat. In other words, they develop a 
toroidal magnetic field component, just like the open field lines 
crossing the outer light cylinder in a pulsar magnetosphere. 
In the language of the Membrane Paradigm (see Thorne~et~al. 1986), 
this toroidal field is needed so that the field lines could slip resistively 
across the stretched event horizon. The next step in our argument 
is to look at those field lines that connect the polar region of
the horizon to the disk somewhere far away from the black hole.
In a force-free magnetosphere, toroidal flux spreads along field 
lines to keep the poloidal current $I\sim B_{\hat{\phi}}\alpha\varpi$ 
constant along the field. Then one can show that the outward pressure
of the toroidal field generated due to the black hole rotation turns 
out to be so large that it cannot be confined by the poloidal field 
tension at large enough distances. In other words, the field lines
under consideration cannot be in a force-free equilibrium. Furthermore, 
one can generalize this argument to the case of closed magnetospheres 
of finite size and derive a conjecture that the maximal radial extent 
$R_{\rm max}$ of the magnetically-coupled region on the disk surface 
should scale inversely with the black hole spin parameter~$a$ in the 
limit $a\rightarrow 0$.

In order to verify this hypothesis and to study the detailed 
structure of magnetically-coupled disk--hole configurations, 
we have obtained numerical solutions of the general-relativistic 
force-free Grad--Shafranov equation corresponding to partially-closed 
field configurations (shown in Fig.~\ref{fig-geometry-kerr}).
This is an nonlinear 2nd-order partial differential equation 
for the poloidal flux function~$\Psi(r,\theta)$ and it is the 
main equation governing the system's behavior.

An additional complication in this problem arises from the need to 
specify two free functions that enter the force-free Grad--Shafranov 
equation; these are the field-line angular velocity $\Omega_F(\Psi)$ 
and the poloidal current~$I(\Psi)$. 
Because all the field lines are assumed to be frozen into 
the disk, the first of this functions is determined in a
fairly straightforward way. Namely, for any given field 
line~$\Psi$, $\Omega_F(\Psi)$ is just the Keplerian 
angular velocity at this line's footpoint on the disk. 
Specifying the poloidal current, on the other hand, is 
a much more difficult and nontrivial task. The reason 
for this is that it cannot be just prescribed 
explicitly on any given surface and one should look 
more thoroughly into the mathematical nature of the 
Grad--Shafranov equation itself to determine~$I(\Psi)$. 
In particular, the most important feature of the Grad--Shafranov 
equation in this regard is that it becomes singular on two surfaces,
the event horizon and the inner light cylinder. This observation is
very useful because one can impose a physically-motivated regularity 
condition at each of these surfaces. One of the most important ideas
in our analysis is that one can use the light-cylinder regularity 
condition to determine, using an iterative procedure, the poloidal
current $I(\Psi)$, similar to the way it was done by CKF99 in the 
context of pulsar magnetospheres. 

As for the singularity at the event horizon, it is also very important. 
Basically, it tells us that it is not possible to prescribe an arbitrary
boundary condition at the horizon; instead, one can only impose a certain
physical condition of regularity there. When combined with the 
Grad--Shafranov equation itself, this regularity requirement results
in a single relationship (historically known as the horizon boundary 
condition, first derived by Znajek 1977) between three functions: the 
horizon flux distribution $\Psi_0(\theta)$, and the two free functions, 
$\Omega_F(\Psi)$ and~$I(\Psi)$ (e.g., Beskin~1997; Beskin \& Kuznetsova~2000). 
What's important is that there are no other independent relationships 
that can be specified on this surface. In practical terms, this means 
that this condition should be used to determine the function 
$\Psi_0(\theta)$ in terms of $\Omega_F(\Psi)$ and~$I(\Psi)$, 
which therefore must be determined outside the horizon. This 
fact helps to alleviate some of the causality issues raised 
by Punsly (1989, 2001, 2003) and by Punsly \& Coroniti (1990).

Since one of the goals of this work was to study the 
dependence $R_{\rm max}(a)$, we performed a series of computations
corresponding to various values of two parameters: the black hole
spin parameter $a$ and the the magnetic link's radial extent $R_s$ 
on the disk surface (the field lines anchored to the disk beyond
$R_s$ were taken to be open and non-rotating). At the same time,
the disk boundary conditions were kept the same in all these runs,
namely, $\Psi_d(r)=\Psi_{\rm tot} r_{\rm in}(a)/r$. Therefore,
varying the value of $R_s$ for fixed $a$ was equivalent to 
varying the amount $\Psi_s$ of open magnetic flux threading 
the disk.

Whereas for some pairs of values of $a$ and $\Psi_s$ we were able 
to achieve a convergent force-free solution, for others we were not. 
Thus, as one of the main results of our computations, we were able to
chart out the allowed and the forbidden domains in the two-parameter
space $(a,\Psi_s)$. The boundary between these two domains is
a curve $a_{\rm max}(\Psi_s)$, which can be easily remapped into
the curve $R_{\rm max}(a)$. As can be seen in Figure~\ref{fig-a_of_Psi_s},
this is a monotonically rising curve with the asymptotic behavior
$a_{\rm max}\propto \Psi_s$ as $\Psi_s\rightarrow 0$, which is in 
line with our predictions.

We also computed the total angular momentum and red-shifted energy exchanged
in a unit of global time~$t$ between the hole and the disk through magnetic 
coupling. We studied the dependence of these quantities on the black hole
spin parameter~$a$ and found that the angular momentum transfer rate rises
roughly linearly with~$a$; it is negative for small~$a$ (meaning the 
angular momentum transfer to the hole) and reverses sign around 
$a\approx 0.23$ (for $\Psi_s=0.5\Psi_{\rm tot}$). The total energy 
transfer increases with~$a$ at an accelerated (i.e., faster than 
linear) rate, especially at larger values of~$a$; it is also negative
at small~$a$, but becomes positive around $a=0.26$. This means that
there is a narrow range $0.23..<a<0.26..$ where the integrated angular
momentum flows from the hole to the disk, whereas the integrated 
red-shifted energy flows in the opposite direction.

Finally, we note that, in the case of open or partially-open
field configuration responsible for the Blandford--Znajek process,
one has to consider magnetic field lines that extend from the event 
horizon out to infinity. Since these field lines are not attached 
to a heavy infinitely conducting disk, their angular velocity 
$\Omega_F(\Psi)$ cannot be explicitly prescribed; it becomes just 
as undetermined as the poloidal current~$I(\Psi)$ they carry. 
Fortunately, however, these field lines now have to cross two 
light cylinders (the inner one and the outer one). Since each 
of these is a singular surface of the Grad--Shafranov equation, 
one can impose corresponding regularity conditions on these two
surfaces. Thus, we propose that one should be able to devise an
iterative scheme that uses the two light-cylinder regularity
conditions in a coordinated manner to determine the two free
functions $\Omega(\Psi)$ and $I(\Psi)$ simultaneously, as a
part of the overall solution process. At the same time, the 
regularity conditions at the event horizon and at infinity 
could be used to obtain the asymptotic poloidal flux distributions 
at $r=r_H$ and at $r\rightarrow\infty$, respectively.
We realise of course that iterating with respect to two
functions simultaneously may be a very difficult task.
This purely technical obstacle (in addition to having to
deal with the separatrix between the open- and closed-field
regions) is the primary reason why, in this paper, we have
restricted ourselves to a configuration which has no open 
field lines extending from the black hole to infinity.
We leave this problem as a topic for future research.

It is possible that, instead of solving the Grad--Shafranov 
equation itself, the easiest and most practical way to achieve 
a stationary solution will be to use a {\it time-dependent} 
relativistic force-free code, such as one of those being
developed now (Komissarov 2001, 2002a, 2004a; MacFadyen 
\& Blandford 2003; Spitkovsky 2004; Krasnopolsky 2004,
private communication).

I am very grateful to V.~Beskin, O.~Blaes, R.~ Blandford, S.~Boldyrev, 
A.~K{\"o}nigl, B.~C.~Low, M.~Lyutikov, A.~MacFadyen, V.~Pariev, 
B.~Punsly, and A.~Spitkovsky for many fruitful and stimulating 
discussions. I also would like to express my gratitude to the 
referee of this paper (Serguei Komissarov) for his very useful 
comments and suggestions that helped improve the paper.
This research was supported by the National Science Foundation 
under Grant No.~PHY99-07949.


\section{References}

Agol, E., \& Krolik, J.~H. 2000, ApJ, 528, 161

Begelman, M.~C., Blandford, R.~D., \& Rees, M.~J. 1984, 
Rev. Mod. Phys., 56, 255

Beskin, V.~S., \& Par'ev, V.~I. 1993, Phys. Uspekhi, 36, 529

Beskin, V.~S. 1997, Phys. Uspekhi, 40, 659

Beskin, V.~S. 2003, Phys. Uspekhi, 46, 1209

Beskin, V.~S., \& Kuznetsova, I.~V. 2000, Nuovo Cimento, 115, 795;
preprint (astro-ph/0004021)

Blandford, R.~D. 1976, MNRAS, 176, 465

Blandford, R.~D. 1999, in Astrophysical Disks:
An EC Summer School, ed. J.~A.~Sellwood \& J.~Goodman 
(San Francisco: ASP), ASP Conf. Ser. 160, 265; preprint 
(astro-ph/9902001)

Blandford, R.~D. 2000, Phil. Trans. R. Soc. Lond. A, 358, 811;
preprint (astro-ph/0001499)

Blandford, R.~D. 2002, in "Lighthouses of the Universe", eds.
Gilfanov, M. et al. (New York: Springer), 381

Blandford, R.~D., \& Znajek, R.~L. 1977, MNRAS, 179, 433 (BZ77)

Blandford, R.~D., \& Payne, D.~G. 1982, MNRAS, 199, 883

Contopoulos, I., Kazanas, D., \& Fendt, C. 1999, ApJ, 511, 351

Damour, T. 1978, Phys. Rev. D, 18, 3589

Fendt, C. 1997, A\&A, 319, 1025

Gammie, C.~F. 1999, ApJ, 522, L57

Ghosh, P., \& Abramowicz, M.~A. 1997, MNRAS, 292, 887

Gruzinov, A. 1999, preprint (astro-ph/9908101)

Hawley, J.~F., \& Krolik, J.~H. 2001, ApJ, 548, 348

Hirose, S., Krolik, J.~H., de~Villiers, J.-P., \& Hawley, J.~F. 2004,
ApJ, 606, 1083

Hirotani, K., Takahashi, M., Nitta, S.-Y., \& Tomimatsu, A. 1992,
ApJ, 386, 455

Komissarov, S.~S. 2001, MNRAS, 326, L41

Komissarov, S.~S. 2002a, MNRAS, 336, 759

Komissarov, S.~S. 2002b, preprint (astro-ph/0211141)

Komissarov, S.~S. 2004a, MNRAS, 350, 427

Komissarov, S.~S. 2004b, MNRAS, 350, 1431

Krolik, J.~H. 1999a, ApJ, 515, L73

Krolik, J.~H. 1999b, Active Galactic Nuclei:
From The Central Black Hole To The Galactic Environment
(Princeton: Princeton Univ. Press)

Levinson, A. 2004, ApJ, 608, 411

Li, L.-X. 2000, ApJ, 533, L115

Li, L.-X. 2001, in X-ray Emission from Accretion onto 
Black Holes, ed. T.~Yaqoob \& J.~H.~Krolik, JHU/LHEA Workshop, 
June 20-23, 2001

Li, L.-X. 2002a, ApJ, 567, 463

Li, L.-X. 2002b, A\&A, 392, 469

Li, L.-X. 2004, preprint (astro-ph/0406353)

Livio, M., Ogilvie, G.~I., \& Pringle, J.~E. 1999, ApJ, 512, 100

Lovelace, R.~V.~E. 1976, Nat., 262, 649

Macdonald, D., \& Thorne, K.~S. 1982, MNRAS, 198, 345 (MT82) 

Macdonald, D.~A. 1984, MNRAS, 211, 313

Macdonald, D.~A., \& Suen, W.-M. 1985, Phys. Rev. D, 32, 848

MacFadyen, A.~I., \& Blandford, R.~D. 2003, AAS HEAD Meeting 35, 20.16 

Nitta, S.-Y., Takahashi, M., \& Tomimatsu, A. 1991, Phys. Rev. D, 44, 2295

Ogura, J., \& Kojima, Y. 2003, Prog. Theor. Phys., 109, 619

Phinney, E.~S. 1983, in Astrophysical Jets, ed. A.~Ferrari \&
A.~G.Pacholczyk (Dordrecht: Reidel), 201

Punsly, B. 1989, Phys. Rev. D, 40, 3834

Punsly, B. 2001, Black Hole Gravitohydromagnetics (Berlin: Springer)

Punsly, B. 2003, ApJ, 583, 842

Punsly, B. 2004, preprint (astro-ph/0407357)

Punsly, B., \& Coroniti, F.~V. 1990, ApJ, 350, 518

Spitkovsky, A. 2004, in IAU Symp. 218, Young Neutron Stars and Their
Environment, ed. F.~M. Camilo \& B.~M. Gaensler (San Francisco: ASP),
357; preprint (astro-ph/0310731)

Takahashi, M. 2002, ApJ, 570, 264

Thorne, K.~S. 1974, ApJ, 191, 507

Thorne, K.~S., Price, R.~H., \& Macdonald, D.~A. 1986,
Black Holes: The Membrane Paradigm (New Haven: Yale Univ. Press)

Uzdensky, D.~A., K{\"o}nigl, A., \& Litwin, C. 2002, ApJ, 565, 1191

Uzdensky, D.~A., 2002a, ApJ, 572, 432

Uzdensky, D.~A., 2002b, ApJ, 574, 1011

Uzdensky, D.~A., 2003, ApJ, 598, 446

Uzdensky, D.~A., 2004, ApJ, 603, 652

van~Ballegooijen, A.~A. 1994, Space Sci. Rev., 68, 299

van~Putten, M.~H.~P.~M. 1999, Science, 284, 115

van~Putten, M.~H.~P.~M., \& Levinson, A. 2003, ApJ, 584, 937

Wang, D.~X., Xiao, K., \& Lei, W.~H. 2002, MNRAS, 335, 655

Wang, D.-X., Lei, W.~H., \& Ma, R.-Y. 2003a, MNRAS, 342, 851

Wang, D.-X., Ma, R.-Y., Lei, W.-H., \& Yao, G.-Z. 2003b, ApJ, 595, 109

Wang, D.-X., Ma, R.-Y., Lei, W.-H., \& Yao, G.-Z. 2004, ApJ, 601, 1031

Znajek, R.~L. 1977, MNRAS, 179, 457

Znajek, R.~L. 1978, MNRAS, 185, 833






\clearpage

\begin{figure}
\plotone{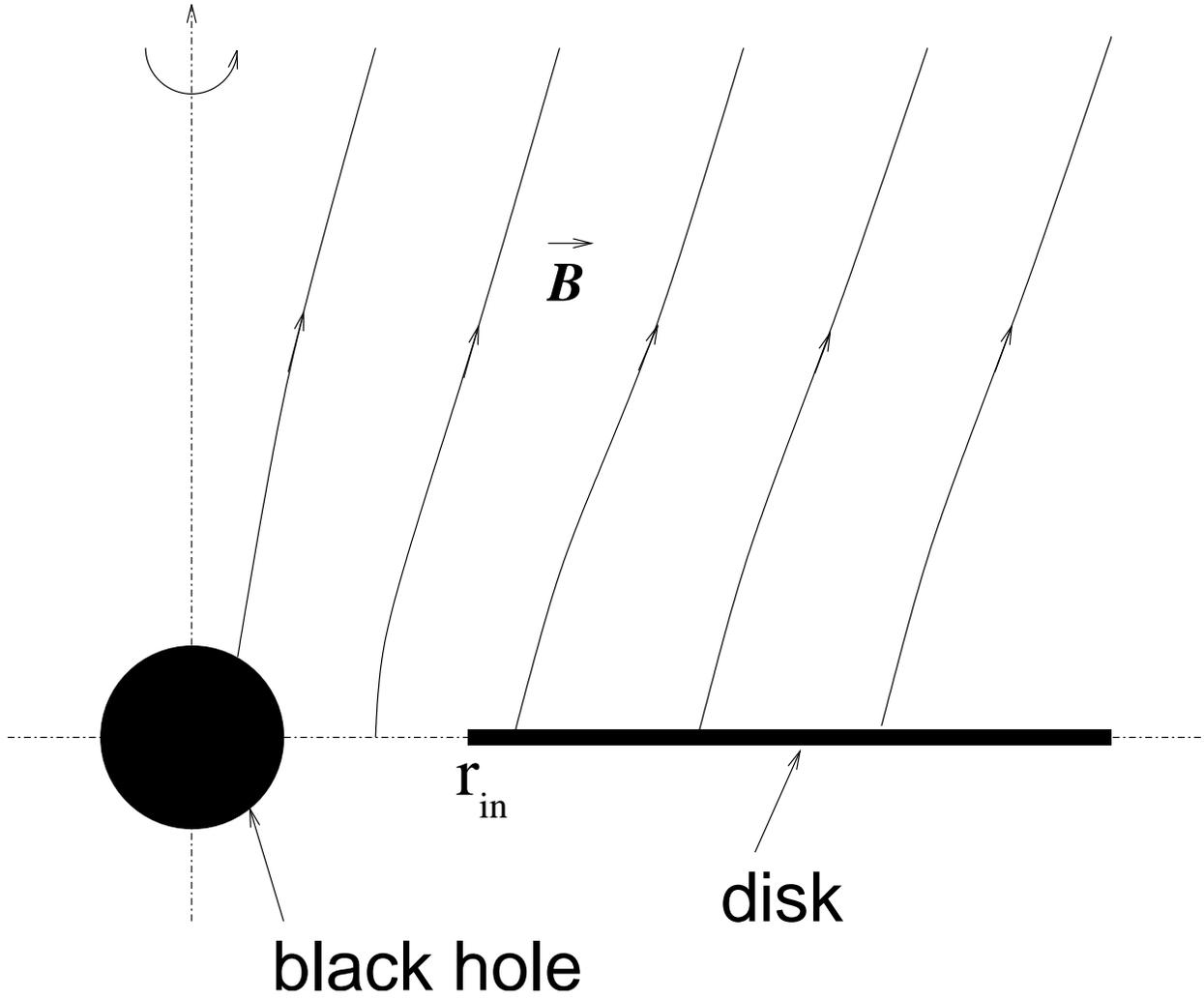}
\figcaption{Schematic drawing of an {\it open} black hole -- disk 
magnetosphere, commonly associated with the Blandford--Znajek (BZ77)
process. Rotational energy and angular momentum are extracted from
both the black hole and the disk and are transported away by the 
magnetic field.
\label{fig-geometry-open}}
\end{figure}

\clearpage

\begin{figure}
\plotone{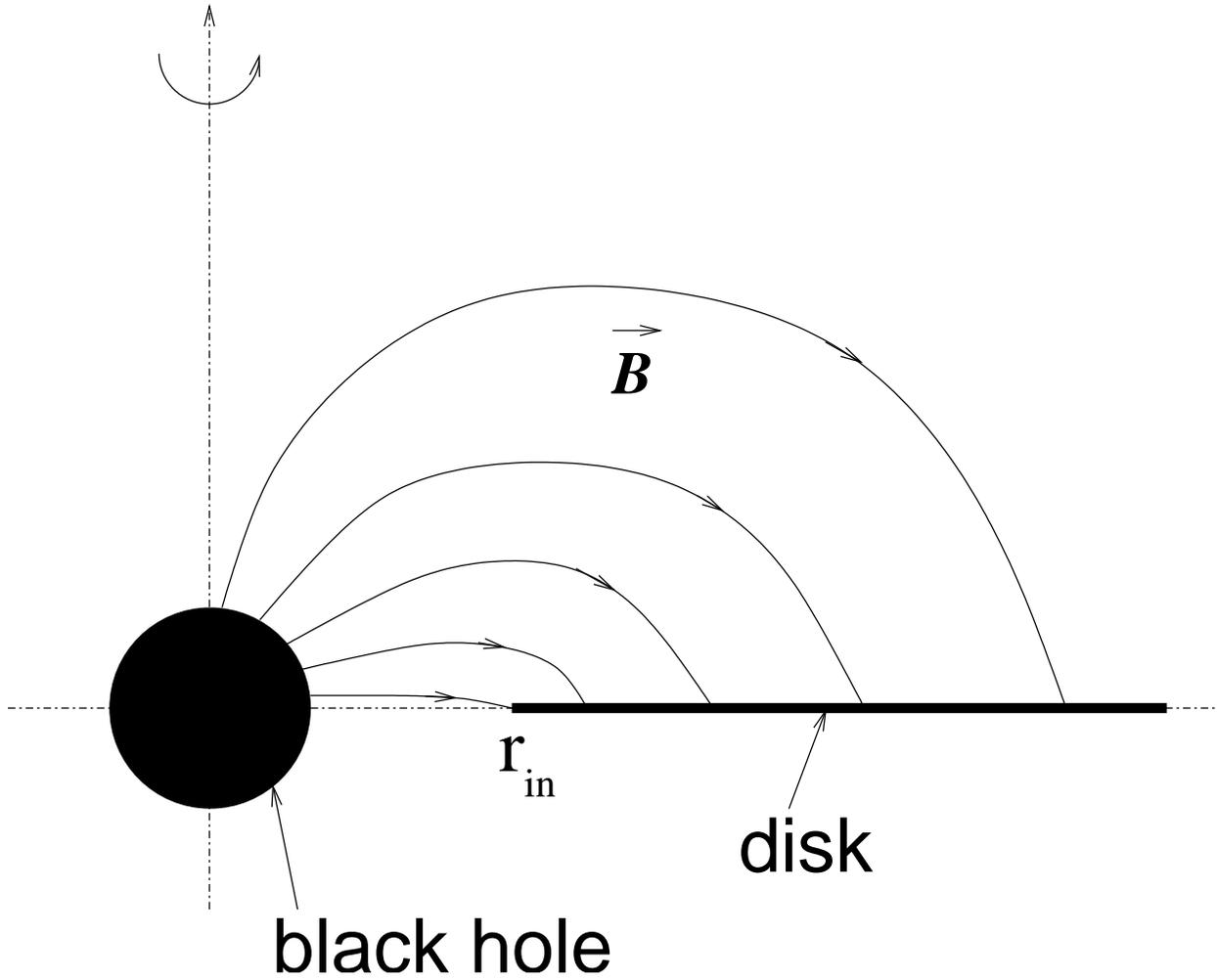}
\figcaption{Schematic drawing of a {\it fully-closed} 
black hole -- disk magnetosphere. Energy and angular 
momentum are exchanged between the hole and the disk
through a direct magnetic link.
\label{fig-geometry-closed}}
\end{figure}

\clearpage

\begin{figure}
\plotone{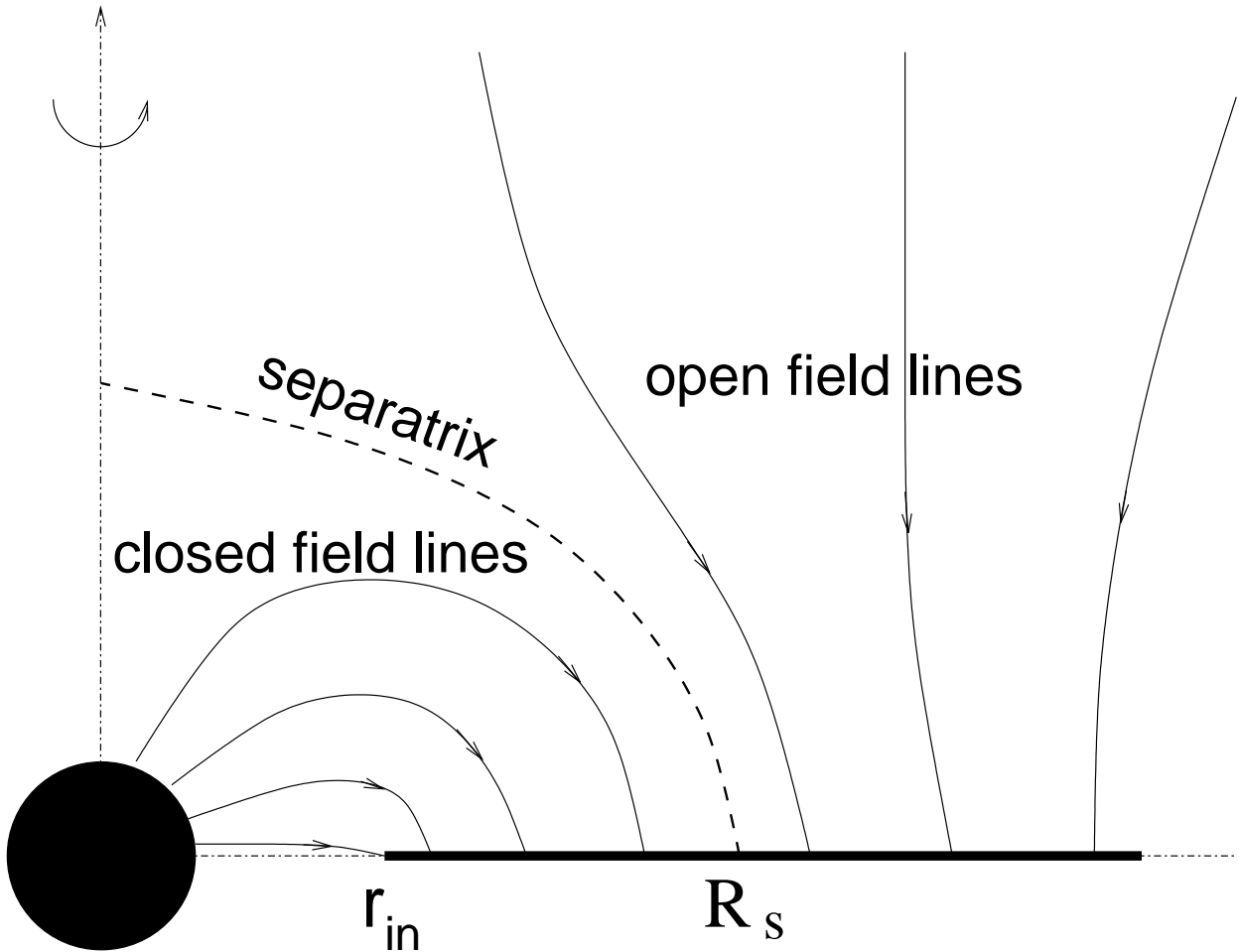}
\figcaption{Schematic drawing of a black hole -- disk magnetosphere
with a {\it radially-limited magnetic connection}. Here, only the
inner part of the disk is coupled magnetically to the hole (closed
field region), whereas the field lines attached to the outer part 
of the disk are open and extend to infinity.
\label{fig-geometry-kerr}}
\end{figure}

\clearpage

\begin{figure}
\plotone{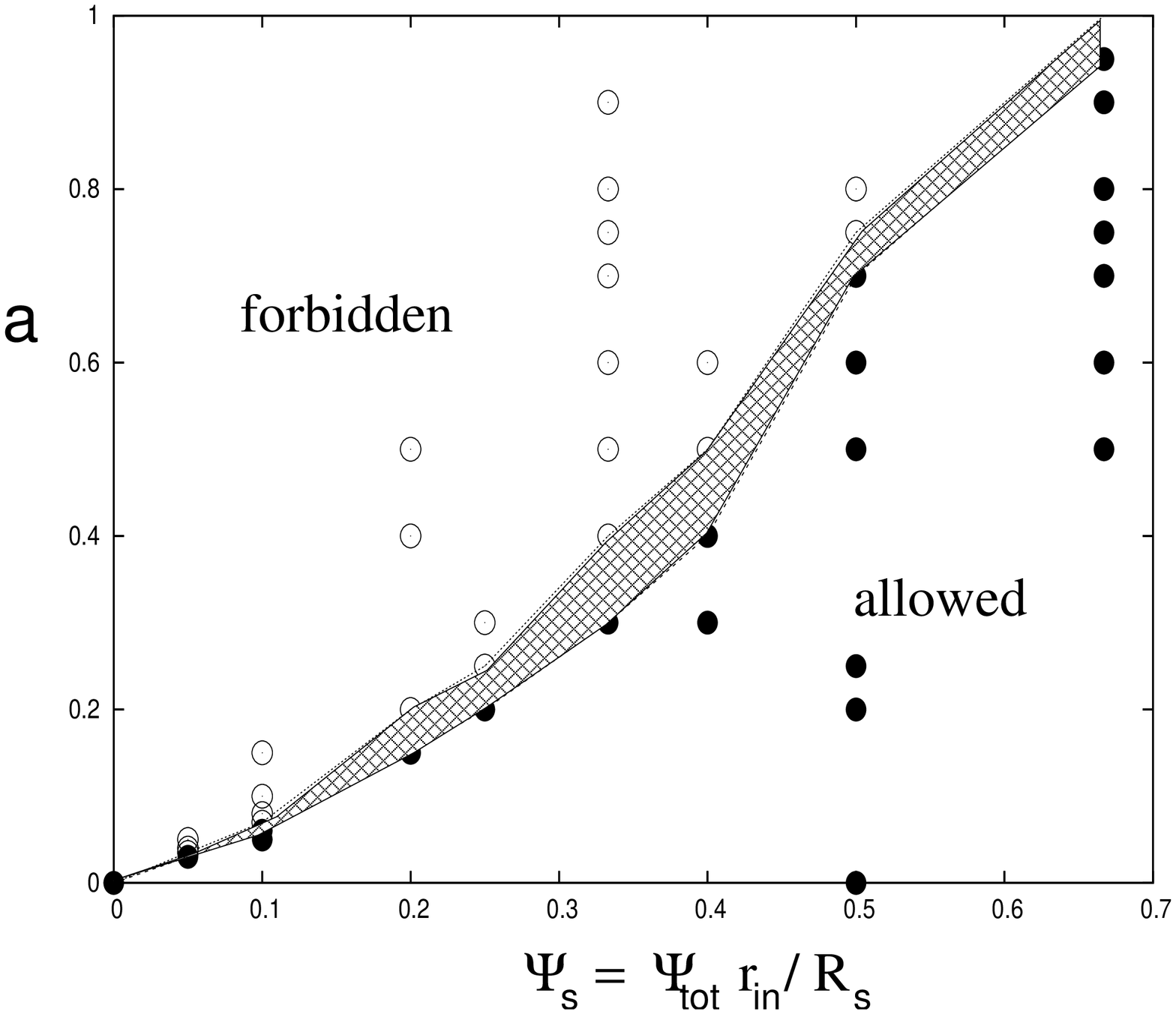}
\figcaption{The 2-D parameter space $(a,\Psi_s)$. Filled (black)
circles correspond to the runs in which a stationary force-free
solution has been obtained, while open (white) circles correspond
to the runs that failed to converge to a stationary solution.
The shaded band running diagonally across the plot represents
the function $a_{\rm max}(R_s)$, the maximal value of $a$ for 
which a force-free magnetic link can extend up to a given radial 
distance $R_s$ on the disk.
\label{fig-a_of_Psi_s}}
\end{figure}

\clearpage

\begin{figure}
\plotone{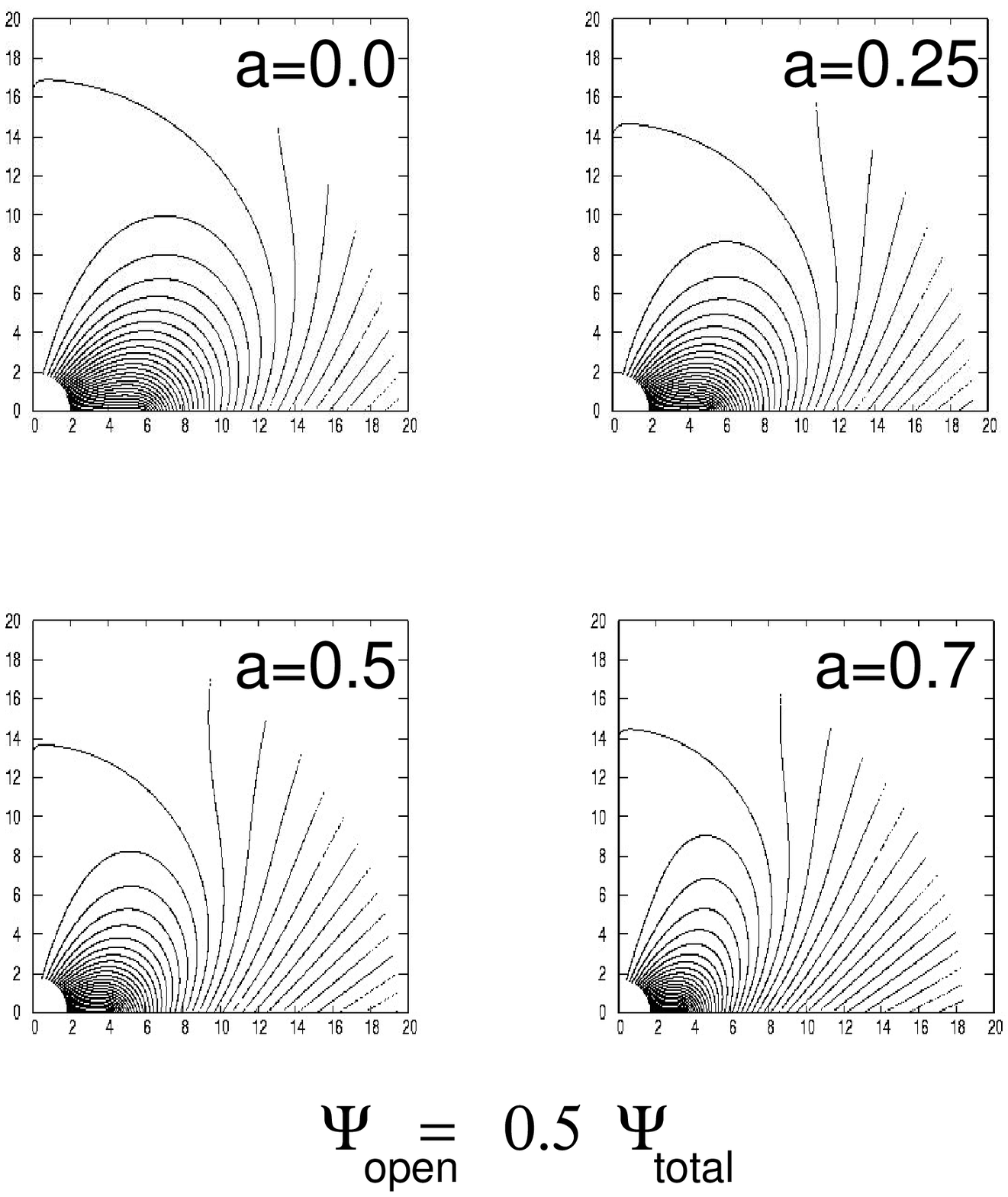}
\figcaption{Contour plots of the magnetic flux function $\Psi(r,\theta)$
for four values of the black hole specific angular momentum: $a=0.0$, 0.25, 
0.5, and~0.7; the amount of open poloidal flux in all cases is 
$\Psi_s=0.5\Psi_{\rm tot}$ (corresponding to $R_s=2r_{\rm in}$). 
\label{fig-contour}}
\end{figure}

\clearpage

\begin{figure}
\plotone{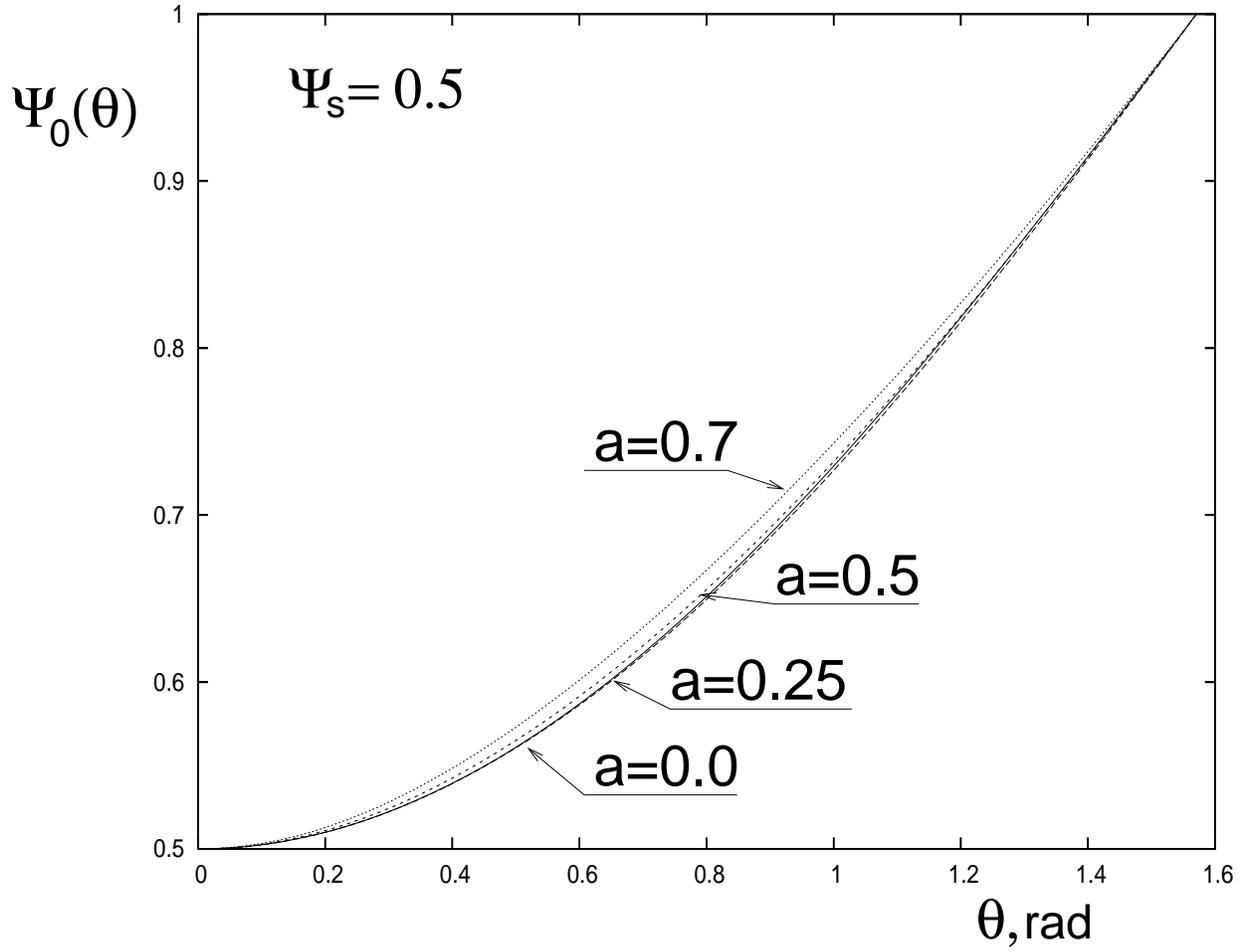}
\figcaption{The horizon poloidal flux distribution $\Psi_0(\theta)$
for several values of $a$ and a single value $\Psi_s=0.5\Psi_{\rm tot}$
of the open magnetic flux (corresponding to $R_s=2\, r_{\rm in}(a)$.
\label{fig-Psi_0}}
\end{figure}

\clearpage

\begin{figure}
\plotone{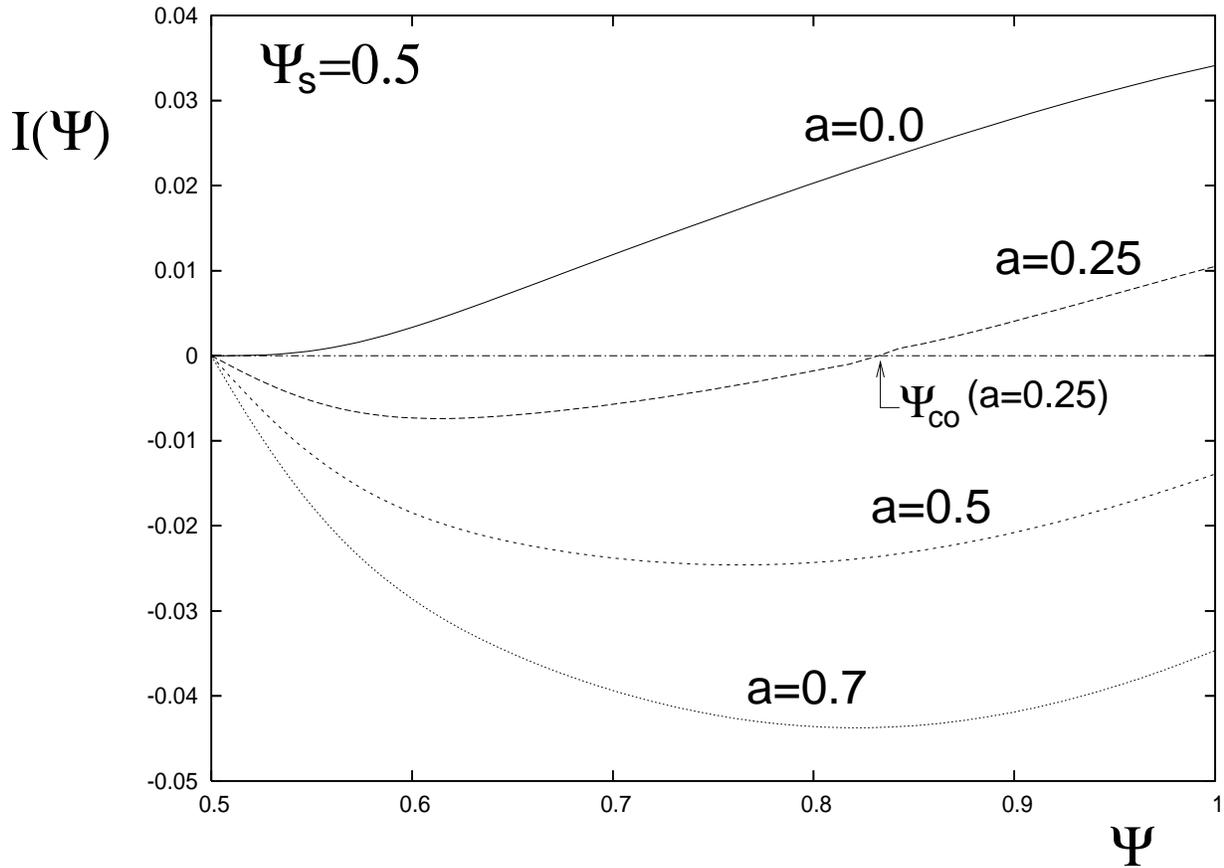}
\figcaption{Poloidal current $I$ as a function of poloidal magnetic 
flux $\Psi$ for several values of $a$ and a single value $\Psi_s=
0.5\Psi_{\rm tot}$.
\label{fig-I}}
\end{figure}

\clearpage

\begin{figure}
\plotone{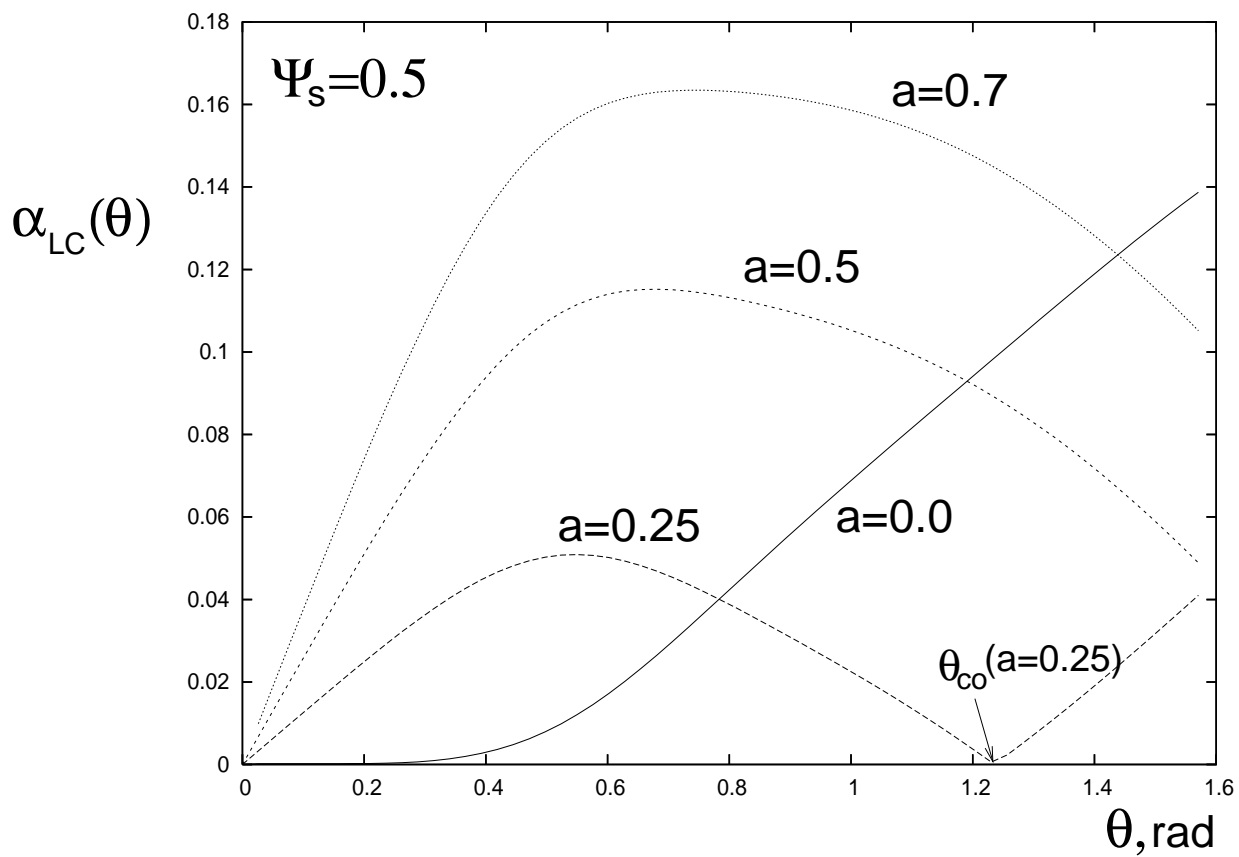}
\figcaption{The position of the inner light cylinder represented
by the lapse function~$\alpha_{\rm LC}(\theta)$ for several values
of~$a$. The light cylinder touches the horizon at the pole $\theta=0$
and, for $a<0.36$, at the point $\theta=\theta_{\rm co}$ where the 
corotation field line $\Psi_{\rm co}\equiv\Psi_d(r_{\rm co})$ 
intersects the horizon.
\label{fig-alpha_LC}}
\end{figure}

\clearpage

\begin{figure}
\plotone{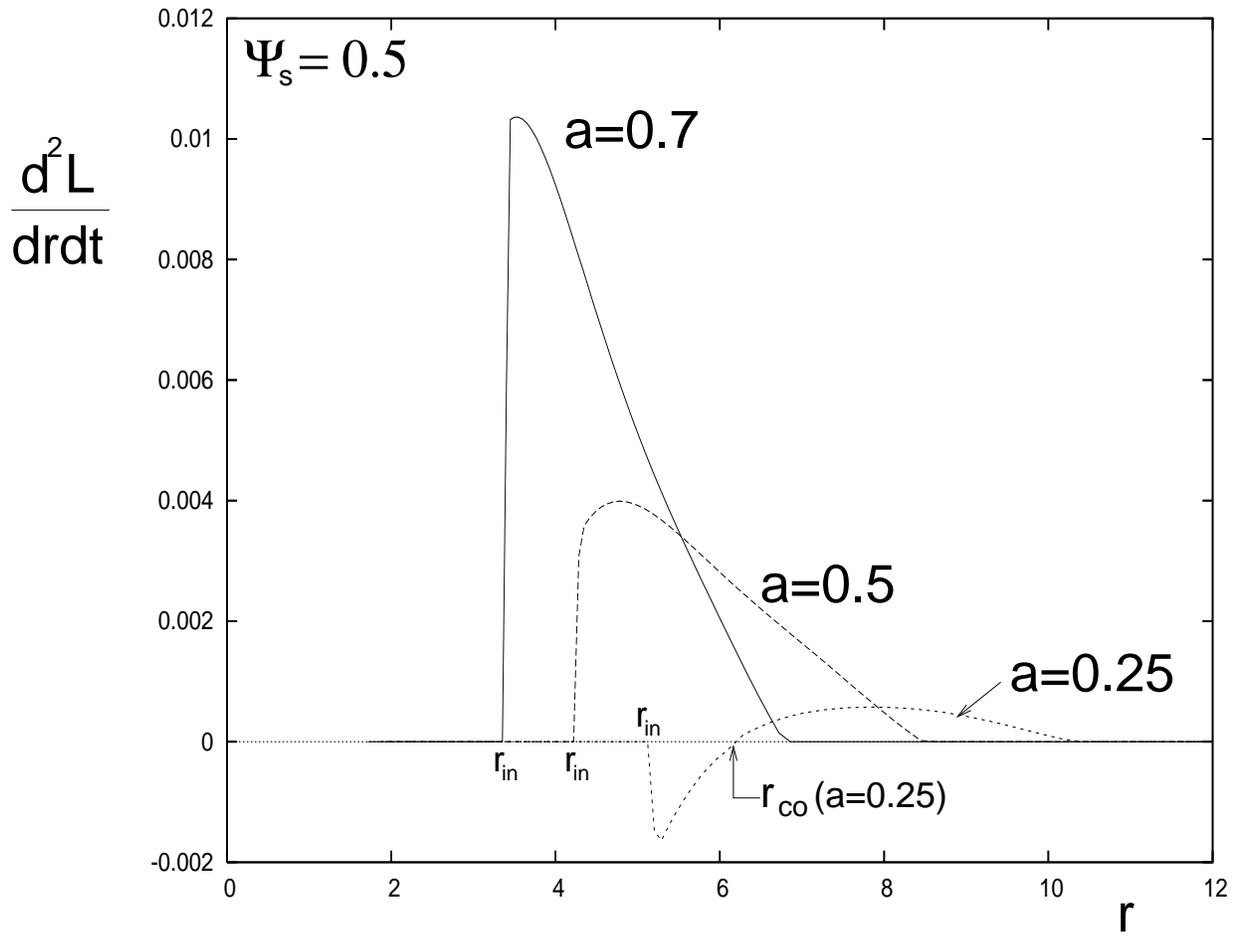}
\figcaption{Radial distribution $d^2L/drdt$ of the magnetic torque 
per unit radius~$r$ on the disk surface for $a=0.25$, 0.5, and~0.7. 
\label{fig-L-of-r}}
\end{figure}

\clearpage

\begin{figure}
\plotone{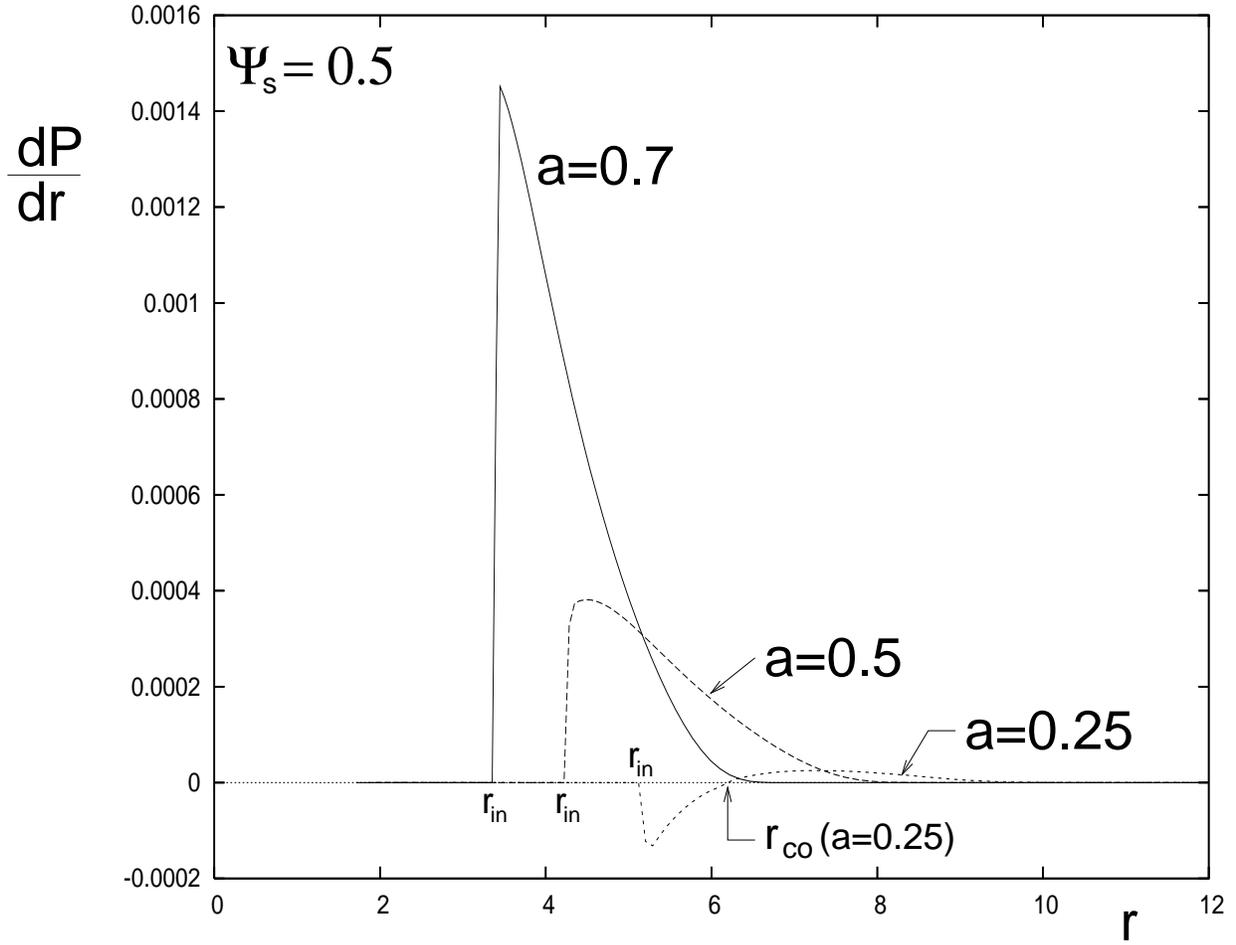}
\figcaption{Radial distribution $dP/dr$ of the red-shifted power 
per unit radius~$r$ on the disk surface for $a=0.25$, 0.5, and~0.7.
\label{fig-P-of-r}}
\end{figure}

\clearpage

\begin{figure}
\plotone{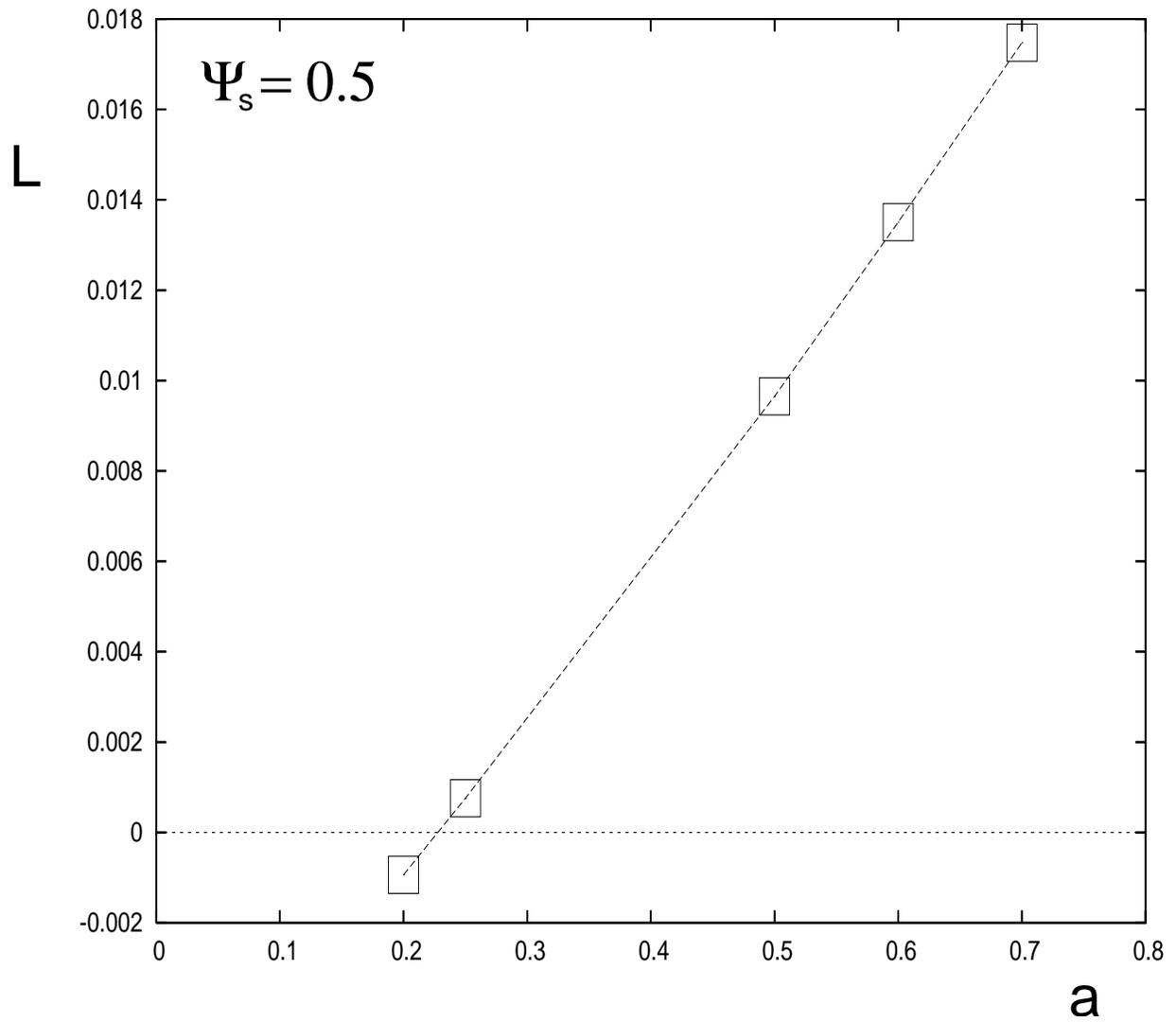}
\figcaption{The dependence of the total magnetic torque between 
the black hole and the disk on the hole's spin parameter~$a$ for
fixed $R_s=2r_{\rm in}$. 
\label{fig-L-of-a}}
\end{figure}

\clearpage

\begin{figure}
\plotone{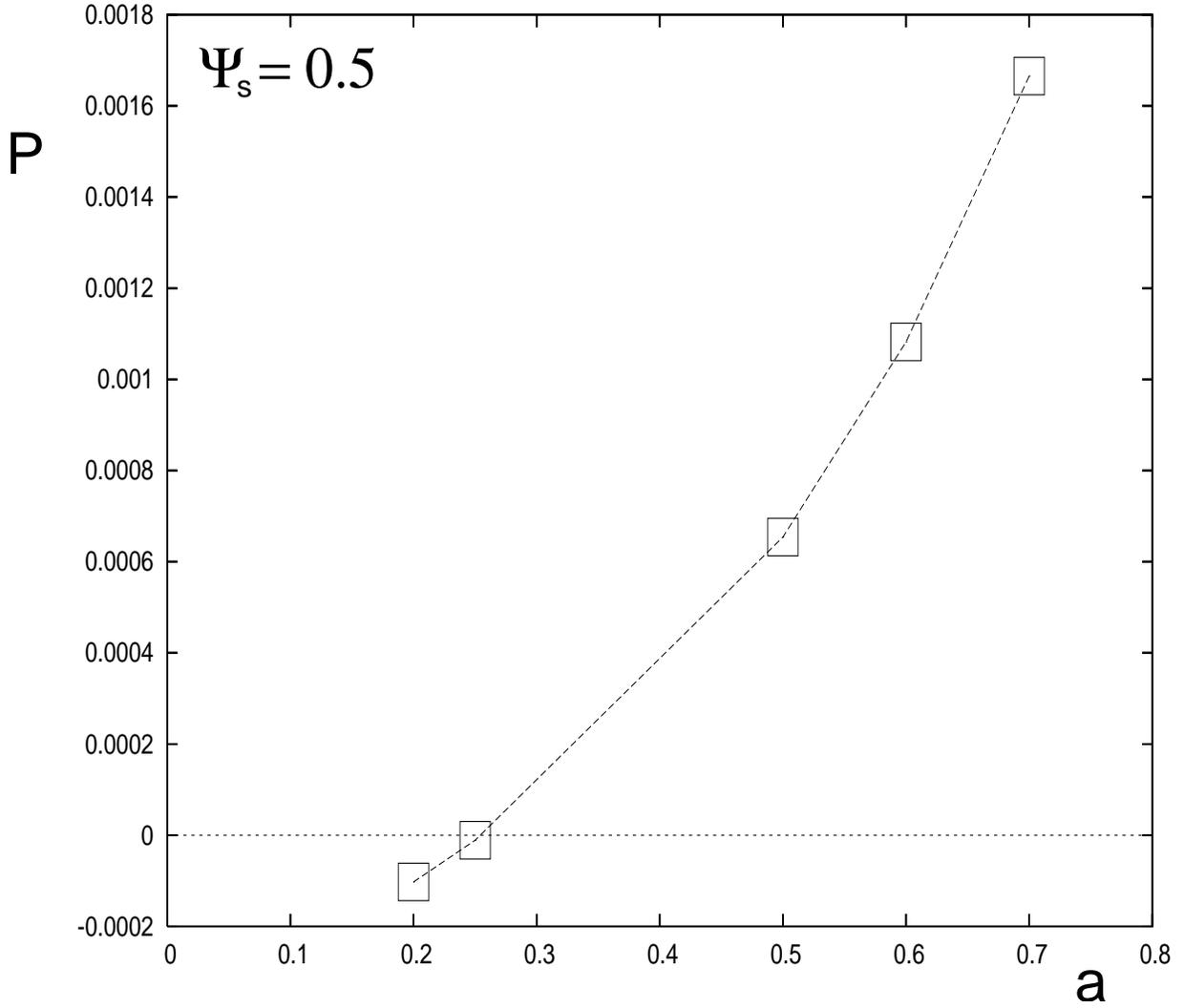}
\figcaption{The dependence of the total red-shifted power exchanged
magnetically between the black hole and the disk on the hole's 
spin parameter~$a$ for fixed $R_s=2r_{\rm in}$.
\label{fig-P-of-a}}
\end{figure}

\end{document}